\documentclass[aps,prd,nofootinbib,superscriptaddress,twocolumn]{revtex4}

\usepackage{amsmath,amsfonts,amssymb}
\usepackage{booktabs}
\usepackage{enumitem}
\usepackage{graphicx}
\usepackage{siunitx}
\usepackage{xcolor}
\usepackage{verbatim}
\usepackage[colorlinks=true,urlcolor=black,citecolor=blue,linkcolor=black]{hyperref}
\usepackage{enumitem}
\setlist[itemize]{leftmargin=*}

\newcommand{\be}{\begin{equation}}
\newcommand{\ee}{\end{equation}}
\newcommand{\bea}{\begin{equation}\begin{aligned}}
\newcommand{\eea}{\end{aligned}\end{equation}}
\newcommand{\td}{{\rm d}}

\newcommand\Hcal{\mathcal{H}}
\newcommand\Lcal{\mathcal{L}}
\newcommand\Ocal{\mathcal{O}}

\newcommand{\R}{\mathbf{R}}

\newcommand{\p}{\mathbf{p}}
\newcommand{\q}{\mathbf{q}}

\newcommand{\TeV}{\rm{TeV}}

\newcommand{\vev}{v_h}
\newcommand{\vel}{v_{\text{rel}}}
\newcommand{\Lambdalin}{\Lambda_{\text{lin}}}
\newcommand{\Lambdaunit}{\Lambda_\star}
\newcommand{\mPl}{m_{\text{Pl}}}

\newcommand{\BR}[1]{\text{BR}_{#1}}

\newcommand{\Lfree}{\Lcal_{\text{free}}}

\newcommand{\TRH}{T_{\text{RH}}}

\newcommand{\parens}[1]{\left(#1\right)}
\newcommand{\brackets}[1]{\left[#1\right]}
\newcommand{\angles}[1]{\left<#1\right>}

\newcommand{\re}[1]{\operatorname{Re}\brackets{#1}}
\newcommand{\im}[1]{\operatorname{Im}\brackets{#1}}

\newcommand{\hc}{\text{h.c.}}

\newcommand{\nicpb}{Laboratory of High Energy and Computational Physics, NICPB, R\"avala pst. 10, 10143 Tallinn, Estonia}

\newcommand{\ippp}{Institute for Particle Physics Phenomenology, Department of Physics, Durham University, Durham DH1 3LE, United Kingdom}

\begin{document}

\title{Dark matter of any spin -- an effective field theory and applications}

\author{Juan C. Criado}
\affiliation{\nicpb}
\affiliation{\ippp}

\author{Niko Koivunen}
\affiliation{\nicpb}
 
\author{Martti Raidal}
\affiliation{\nicpb}

\author{Hardi Veerm\"ae}
\affiliation{\nicpb}

\date{\today}

\begin{abstract}
   We develop an effective field theory of a generic massive particle of any spin and, as an example, apply this to study higher-spin dark matter (DM). Our formalism does not introduce unphysical degrees of freedom, thus avoiding the potential inconsistencies that may appear in other field-theoretical descriptions of higher spin. Being a useful reformulation of the Weinberg's original idea, the proposed effective field theory allows for consistent computations of physical observables for general-spin particles, although it does not admit a Lagrangian description. As a specific realization, we explore the phenomenology of a general-spin singlet with $\mathbb{Z}_2$-symmetric Higgs portal couplings, a setup which automatically arises for high spin, and show that higher spin particles with masses above $O(10)\,\TeV$ can be viable thermally produced DM candidates. Most importantly, if the general-spin DM has purely parity-odd couplings, it naturally avoids all DM direct detection bounds, in which case its mass can lie below the electroweak scale. Our formalism reproduces the existing results for low-spin DM, and allows one to develop consistent higher-spin particle physics phenomenology for high- and low-energy experiments and cosmology.
\end{abstract}

\maketitle

\section{Introduction}
\label{sec:intro}
\enlargethispage{\baselineskip}

By very general considerations~\cite{Weinberg:1964ew,Grisaru:1977kk,Grisaru:1976vm,Weinberg:1980kq,Porrati:2008rm}, massless interacting fundamental constituents of matter can have at most spin 2. Dark matter (DM), being necessarily massive, naturally avoids conditions of those theorems and can thus consist of particles with arbitrary spin. 

Spin-0~\cite{Chatrchyan:2012ufa,Aad:2012tfa}, spin-1/2, spin-1 and, as manifested by the discovery of gravitational waves~\cite{Abbott:2016blz}, also spin-2 particles do exist in nature. A generic pattern appears in the conventional description of these particles: the higher the spin of the particle, the more constraints are needed to describe it. The first example is fundamental massive vector bosons, which must be gauge bosons.

For higher spins the situation becomes more complicated. For example, motivated by Dirac's successful description of relativistic spin-1/2 fermions~\cite{Dirac:1936tg}, and the Fierz-Pauli theory of particles with general spin~\cite{Fierz:1939ix}, Rarita and Schwinger proposed a first-order derivative theory of a generic spin-3/2 field~\cite{Rarita:1941mf}. However, the massive Rarita-Schwinger field contains unphysical degrees of freedom which must be projected out. Generically, in interacting theories the eliminated degrees of freedom reappear, causing potential pathologies including the violation of causality and perturbative unitarity~\cite{Johnson:1960vt,Velo:1970ur}. The only known consistent way to get rid of the unphysical background field is to embed the Rarita-Schwinger theory in supergravity~\cite{Freedman:1976xh, Deser:1976eh, Freedman:1976py,VanNieuwenhuizen:1981ae}. This implies that the spin-3/2 field must be identified with the gravitino~\cite{Grisaru:1977kk,Grisaru:1976vm}. 
Needless to say, this also implies that the low-energy limit of this theory contains other fields in addition to the gravitino, the superpartners, and that the couplings of all those particles are fixed by supersymmetry.

In general, in the absence of ultraviolet (UV)-complete theory of higher-spin fields, the questions of consistency and physical viability will always arise whenever one computes any physical observable involving higher-spin degrees of freedom. At the same time, there is considerable interest to the phenomenology of higher-spin fields. Higher-spin resonances are known to exist in nuclear physics: thus, one must be able to compute their cross sections in order to interpret experimental results. Higher-spin particles can also form the DM of the Universe, which implies that one must be able to compute their freeze-in or freeze-out cross sections and low-energy interactions with the Standard Model (SM) matter in direct detection experiments. As higher-spin particles appear in the extensions of gravity and supergravity, they can appear as resonances at high-energy colliders or as a specific fifth force in low-energy experiments. In all those cases, the crucial question stands -- is there a consistent framework in which physical observables for higher-spin particles can be computed? 

By far the most studied higher-spin particle is the spin-3/2 fermion (spin-2 DM has also been studied in~\cite{Aoki:2016zgp,Babichev:2016hir,Babichev:2016bxi,Marzola:2017lbt}). Leaving aside the gravitino phenomenology, attempts to describe a spin-3/2 fermion with generic interactions to the SM are dominated by studies of the Rarita-Schwinger field~\cite{Kamenik:2011vy,Yu:2011by,Ding:2012sm,Ding:2013nvx,Dutta:2015ega,Khojali:2016pvu,Khojali:2017tuv,Chang:2017dvm,Goyal:2019vsw,Garcia:2020hyo}. Unfortunately, as explained above, the interacting Rarita-Schwinger field in the Lorentz representation $(1, 1/2)$ contains unphysical degrees of freedom. In nuclear physics, where spin-3/2 resonances must be described, physicists avoid the extra degrees of freedom by using second-derivative Lagrangians with specifically chosen interactions to describe the $(3/2, 0)$ representation~\cite{Delgado-Acosta:2013kva,DelgadoAcosta:2015ypa,Mart:2019jtb}. We argue that, in this case, even the free field theory cannot be quantized consistently without introducing new fields and constraints (see Appendix \ref{app:quantization-fermions}), despite the fact that the SM does admit the second-order formulation~\cite{Chalmers:1997ui,Dreiner:2008tw}.
An effective field theory (EFT) for  any integer-spin particle has been constructed in Ref.~\cite{Bellazzini:2019bzh} using $(j/2, j/2)$ fields. However, these fields contain unphysical components.\footnote{It has been suggested that issues related to unphysical modes might be treated by removing them from the spectrum of asymptotic states~\cite{Anselmi:2020opi}.} 
Is it possible to have an EFT for interacting particles of general spin without introducing such extra degrees of freedom? The answer is yes.

In this work we develop an EFT describing a generic massive particle of any spin in which only the physical degrees of freedom are introduced from the very beginning. The idea goes back to Weinberg~\cite{Weinberg:1964cn}, who uses a non-Lagrangian field theory with fields in the $(j, 0)$ representation to compute scattering amplitudes for physical processes. Unfortunately, in Weinberg's notation, the formalism becomes increasingly more complicated the higher the particle's spin is, and we are not aware of any practical computation performed using this formalism. 

Here, we propose an EFT framework realizing this idea which easily allows for computations and remains unchanged for any spin (see Appendix~\ref{app:notation} for the notation). This EFT does not admit a Lagrangian description. Needless to say, problems related to the presence of unphysical components, such as the violation of causality, are absent. On the other hand, perturbative unitarity is unavoidably broken in our description at some high scale $\Lambda$ above the particle's mass. Thus our proposal represents an effective tool for physically meaningful and consistent computations of higher-spin particle observables with generic couplings to the SM fields. Unlike in supersymmetry, no relation between different couplings needs to be imposed.

We do not consider gravitational interactions. However, general arguments in string theory indicate that additional light particles in the gravity sector must exist if higher-spin particles exist in nature~\cite{Afkhami-Jeddi:2018apj,Kaplan:2020ldi}. These arguments further motivate studies of higher-spin particles.

As an application of our framework, we show that such a particle can provide a natural DM candidate, when stabilized by a $\mathbb{Z}_2$ symmetry under which only the higher-spin particle is odd, while all the SM ones are even. For low spins this symmetry has to be imposed by hand, but when the spin is sufficiently high, an accidental $\mathbb{Z}_2$ symmetry is naturally realized. Indeed, in order for an interaction to explicitly break this symmetry, it must contain the higher-spin particle an odd number of times. Thus, the higher the spin, the more SM particles are needed to construct Lorentz-invariant local operators. The effects of such an interaction will be suppressed by powers of $m/\Lambda$, where $m$ is the mass of higher-spin particle. Making either $\Lambda$ or $j$ large will then render the higher-spin particle metastable.

We work out the DM results for the lowest order coupling of a general-spin particle, the Higgs portal, and demonstrate that the observed DM abundance can be obtained both for freeze-in and freeze-out processes, consistently with the known results in the case of low-spin particles. A particularly important result concerns DM direct detection -- for purely $P$-odd couplings the direct detection cross section is naturally suppressed, providing a possible explanation to the nonobservation of higher-spin DM in those experiments. 

More generally, our results enable one to work consistently with generic higher-spin fields and to develop phenomenology of those particles without worrying about the possible disastrous effects from unphysical degrees of freedom.

This paper is organized as follows.  In Sec.~\ref{sec:EFT} we develop the EFT of general-spin particles. Sec.~\ref{sec:pheno} deals with the phenomenology of general-spin DM from the Higgs portal. Our main results are presented in Sec.~\ref{sec:results}, and we conclude in Sec.~\ref{sec:conclusions}. Various technical results are presented in the Appendixes.  In particular, we introduce our notation in Appendix~\ref{app:notation}, we demonstrate in Appendix~\ref{app:quantization-fermions} that quantization of second order fermions without additional fields and constraints is inconsistent, we compute  corrections to propagators in our framework and discuss their interpretation in Appendix~\ref{app:quadratic}, we collect several alternatives of spin-3/2 field beyond the minimal representation in Appendix~\ref{app:spin-32}, and we present results of cross section computations in Appendix~\ref{app:xs-formulas}. Throughout the paper we use natural units $\hbar = c = 1$ and the metric signature $(+,\!-,\!-,\!-)$.

\section{Effective field theory for general-spin particles}
\label{sec:EFT}

The choice of a field that describes a particle of any mass and spin is not unique. In particular, the creation and annihilation operators for a massive particle of spin $j$ may be contained in any field transforming as the $(l, r)$ irrep of the Lorentz group, whenever $|l - r| \leq j \leq l + r$ and $l + r + j$ is an integer number. However, not all possibilities have equal practical importance. The irreps $(j, 0)$ and $(0, j)$ are minimal in the sense that they contain exactly the necessary number of degrees of freedom for this purpose. In contrast, any other irrep will include unphysical components, making the construction of a theory for them more complicated. In this case additional care is needed not to couple these components with any physical degree of freedom.\footnote{Despite this, nonminimal irreps can be useful in specific cases, as it happens for the usual vector $(1/2, 1/2)$ fields, and for the Rarita-Schwinger $(1, 1/2) \oplus (1/2, 1)$ field.}
 In this work we will make use of the minimal irreps only. 

Implementing parity transformations, the field space requires, in principle, existence of  a pair of fields $\psi_L \sim (j, 0)$ and $\psi_R \sim (0, j)$. One can collect both in a multiplet $(\psi_L, \psi_R)$ belonging to the representation $(j, 0) \oplus (0, j)$. This is the minimal field content for charged particles. However, if all the symmetries act on the particle through real representations, a single field $\psi$ with $j$ components, known as a \emph{purely neutral} field, is needed. The representation $\R_j$ to which this field belongs to is defined as the subspace of $(j, 0) \oplus (0, j)$ for which $\psi_L^\dagger = \psi_R$. We will focus on purely neutral fields in the rest of this work, which constitute the minimal option for describing DM, since the DM particles themselves are neutral. We will call these fields \emph{general-spin fields}. Using the example of spin-$3/2$, a brief comparison of minimal and nonminimal possibilities is given in Appendix~\ref{app:spin-32}.

\subsection{Free theory}
\label{sec:free}

In order to outline the perturbative setup, we must begin with the basic ingredients of a free theory of massive general-spin particles. Our fields will be in the $\R_j$ representation. Similar arguments and conclusions as the ones presented here apply to the charged version of this representation. The Feynman rules for such fields were derived by Weinberg more than half a century ago~\cite{Weinberg:1964cn}. No Lagrangian formulation for the free sector of this theory is known. Below, we outline some of the theoretical difficulties one encounters in attempting to construct such a formulation. They strongly suggest that it does not exist. Once this is shown, we will adopt Weinberg's non-Lagrangian field-theoretical approach~\cite{Weinberg:1964cn}, whose Feynman rules take a very simple form in our symmetric multispinor index notation.

The free $\R_j$ field can be decomposed in terms of the creation and annihilation operators of the one-particle states:
\begin{align}
    \psi_{(a)}(x)
    =
    \int \frac{\td^3p}{(2\pi)^3 (2 E_\p)}
    \sum_\sigma \Big[
       & a_{p\sigma} \, u_{(a)}(p, \sigma) e^{i p x} \nonumber \\[5pt]
        +& \; a^*_{p\sigma} \, v_{(a)}(p, \sigma) e^{-i p x}\Big]
    ,
    \label{eq:expansion}
\end{align}
where $E_\p^2 = \p^2 + m^2$, $p = (E_\p, \p)$, and $(a) \equiv a_1 \ldots a_{2j}$ is a symmetrized multi-index built from two-component spinor indices (for details, see Appendix~\ref{app:notation}). The creation operators $a$, $a^*$ satisfy the following (anti)commutation relations:
\begin{align}
    [a_{p\sigma}, a^*_{q\rho}]_\pm 
    =
    (2\pi)^3 (2 E_\p) \delta_{\sigma\rho} \delta^3(\p - \q),
    \label{eq:ccr}
\end{align}
with $[\cdot,\cdot]_\pm$ being the commutator for bosons and the anticommutator for fermions. The action of the Poincar\'e group over the creation and annihilation operators can be determined from its action over the one-particle states. The wave functions $u$ and $v$ are completely determined by the transformation properties of $\psi$ and $a$. For our current purposes, it is sufficient to state the explicit expressions for the spin sums of $u$ and $v$, which are shown in Ref.~\cite{Weinberg:1964cn} to be:
\begin{align}
    \sum_\sigma u_{(a)}(p, \sigma) u^*_{(\dot{a})}(p, \sigma)
    &= \frac{p_{(a)(\dot{a})}}{m^{2j}},
    \label{eq:completeness-1}
    \\
    \sum_\sigma v_{(a)}(p, \sigma) v^*_{(\dot{a})}(p, \sigma)
    &= \frac{p_{(a)(\dot{a})}}{m^{2j}},
    \label{eq:completeness-2}
    \\
    \sum_\sigma u_{(a)}(p, \sigma) v^{(b)}(p, \sigma)
    &= {\delta_{(a)}}^{(b)},
\end{align}
where $p_{(a)(\dot{a})} \equiv p_{a_1\dot{a}_1} \ldots p_{a_{2j}\dot{a}_{2j}}$, $\delta_{(a)}^{(b)} \equiv \delta_{a_1}^{b_1} \ldots \delta_{a_{2j}}^{b_{2j}}$, and symmetrization over all indices of the same type at the same height is implied. We remark that these equations are considerably simpler in our notation than in the spin-index notation used in Ref.~\cite{Weinberg:1964cn}.

The field $\psi$ has mass dimension one regardless of the spin, as is inferred from the normalization of $a$, $a^{*}$ and $u_{(a)}$, determined by Eq.~\eqref{eq:ccr}, and Eqs.~\eqref{eq:completeness-1},\eqref{eq:completeness-2}), respectively. This leads to an unconventional mass dimension already for spin $j=1/2$. The conventional case is recovered by rescaling the wave functions with $m^{j}$. Such a rescaling will, however, modify the mass dimension of the coupling. To remove any arbitrariness related to conventions, we will consider the higher-spin field $\psi$ to be effectively
\be
    \Delta_\psi \equiv 1 + j
\ee   
dimensional. This choice is justified later when considering interactions.

From the Lorentz-group transformation properties of $u$ and $v$, it follows that the fields satisfy the following order-$2j$ equations~\cite{Weinberg:1964cn}:
\begin{align}
    \partial^{(\dot{a})(a)} \psi_{(a)} &= m^{2j} \psi^{\dagger \, (\dot{a})},
    \label{eq:eom}
\end{align}
where $\partial^{(\dot{a})(a)} \equiv \partial^{\dot a_1 a_1} \ldots \partial^{\dot a_{2j} a_{2j}}$, with symmetrization over all indices of the same type is to be understood. In addition, as with any other relativistic field, all their components must satisfy the Klein-Gordon equation:
\begin{align}
    (\square + m^2) \psi_{(a)} &= 0.
    \label{eq:klein-gordon}   
\end{align}

The difficulties with constructing a free Lagrangian for these field equations becomes apparent when noting that the number of field equations \eqref{eq:eom} and \eqref{eq:klein-gordon} is twice as large as the number of field components. Thus, in a Lagrangian theory, not all of the field equations can be independent or the additional equations must appear as constraints, either hidden or explicitly imposed.\footnote{One could try for example the Lagrangian
\begin{align}\label{fn:lag}
    \Lfree &=
    \kappa_1 \parens{
        \psi^{(a)} \partial_{(a)(\dot{a})} \psi^{\dagger \, (\dot{a})}
        - \frac{m^{2j}}{2} (\psi^{(a)} \psi_{(a)} + \hc),
    }
    \nonumber \\
    &\phantom{=}
    + \kappa_2 \parens{
        \psi^{(a)} (\square + m^2) \psi_{(a)} + \hc
    },
    \nonumber
\end{align}
which gives Eq.~\eqref{eq:eom} for $\kappa_1 \neq 0 = \kappa_2$, and Eq.~\eqref{eq:klein-gordon} for $\kappa_1 = 0 \neq \kappa_2$, but it cannot produce both for constant parameters $\kappa_1$ and $\kappa_2$.} For scalars, $j=0$, this issue is solved trivially as only the Klein-Gordon equation applies in this case. For $j=1/2$, the Klein-Gordon equation can be derived from the Majorana equation~\eqref{eq:eom}. However, for $j>1$, Eq.~\eqref{eq:eom} is of higher order than the Klein-Gordon equation~\eqref{eq:klein-gordon}. One cannot derive the former from the latter, since the former mixes different components of the fields, and the latter consists only of one independent equation for each component.

\begin{figure}
    \centering
    \includegraphics[width=0.3\textwidth]{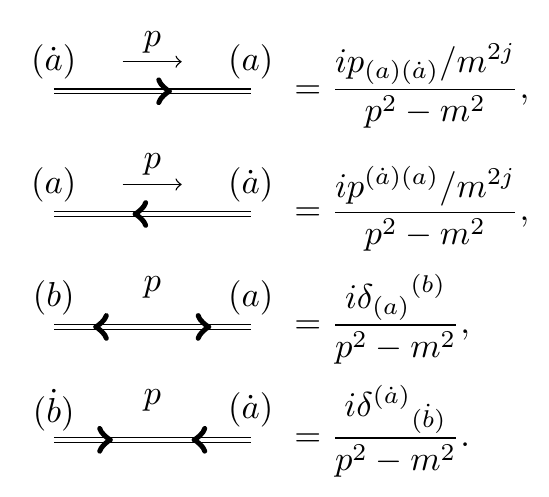}
    \caption{Feynman rules for internal lines in the $\R_j$ theory.}
    \label{tab:propagators}
\end{figure}

Obstructions to the construction of a free Lagrangian manifest also in other ways. For example, one expects the Lagrangian to be of the schematic form $\psi P^{-1} \psi$, where $P$ is the free propagator. From Eq.~\eqref{eq:expansion} one can compute $P$, which is given by the Feynman rules in Fig.~\ref{tab:propagators}. Inverting these functions would give a nonlocal Lagrangian. Additional but unrelated problems appear in fermionic theories, since Lorentz invariance requires that the kinetic terms for $j>1/2$ have more than one derivative, and this leads to complications (see Appendix~\ref{app:quantization-fermions}). Finally, for $j>1$, quadratic operators $\mu^2\psi^{(a)}\psi_{(a)} + \hc$ are not simple mass terms since they can generate additional poles to the propagator at high scales (see Appendix~\ref{app:quadratic}).

These facts strongly suggest that a consistent Lagrangian formulation for the theory of purely $\R_j$ quantum fields does not exist. If it requires the introduction of extra degrees of freedom and constraints, the main advantage of the $(j, 0)$ representation, which is that they only contain physical degrees of freedom, would be lost. Since the Feynman rules for propagators and external legs can be computed without relying on a Lagrangian~\cite{Weinberg:1964cn}, we will abandon the Lagrangian formulation for the rest of this work.

\begin{figure}
    \centering
    \includegraphics[width=0.45\textwidth]{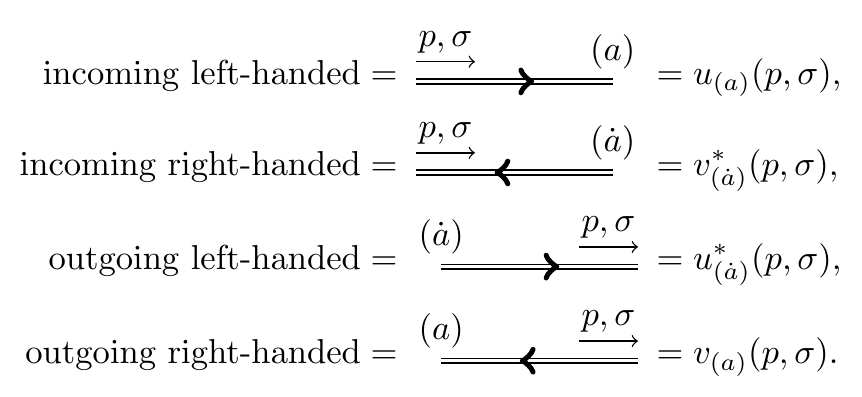}
    \caption{Feynman rules for external lines in the $\R_j$ theory. $\sigma = j, j - 1, \ldots, -j$ denotes the spin state of the external particle.}
    \label{tab:external-legs}
\end{figure}

The propagators and external lines corresponding to the field equations \eqref{eq:eom} and \eqref{eq:klein-gordon} are shown in Figs.~\ref{tab:propagators} and~\ref{tab:external-legs}.\footnote{We remark that the omnipresent object $\Pi(p)$ in Ref.~\cite{Weinberg:1964cn} is now simply $p^{(a)(\dot{a})}$. The equivalence between $\Pi(p)$ and $p^{(a)(\dot{a})}$ follows from the fact that there is only one object with the required transformation properties and that they are normalized in the same way.} We stress that the initial and final states are $u_{(a)} \oplus v^*_{(\dot{a})}$ and $u^*_{(\dot{a})} \oplus v_{(a)}$, respectively. Technically this means that amplitudes must be built from diagrams with \emph{all possible} orientations of the external legs. In the fermionic case, the usual sign rules arising from permutations of the external legs and loops apply.

\subsection{Interactions}
\label{sec:interactions}

The Hamiltonian density $\Hcal_{\text{free}}$ for the free theory defined through the Feynman rules above can be perturbed by adding an interacting Hamiltonian density $\Hcal_{\text{int}}$. One can then derive Feynman rules for the interactions introduced through $\Hcal_{\text{int}}$, defining in this way a perturbative theory with local interactions. Our assumption is that below some energy scale $\Lambda$, much larger than the mass $m$ of the particle and the electroweak scale, the only degrees of freedom present are those of the SM together with the general-spin particle. The situation is then describable by an EFT in which the effects of the new physics at $\Lambda$ are incorporated through nonrenormalizable local operators whose effects are suppressed by inverse powers of $\Lambda$.

\begin{figure*}
    \centering
    \includegraphics[width=\textwidth]{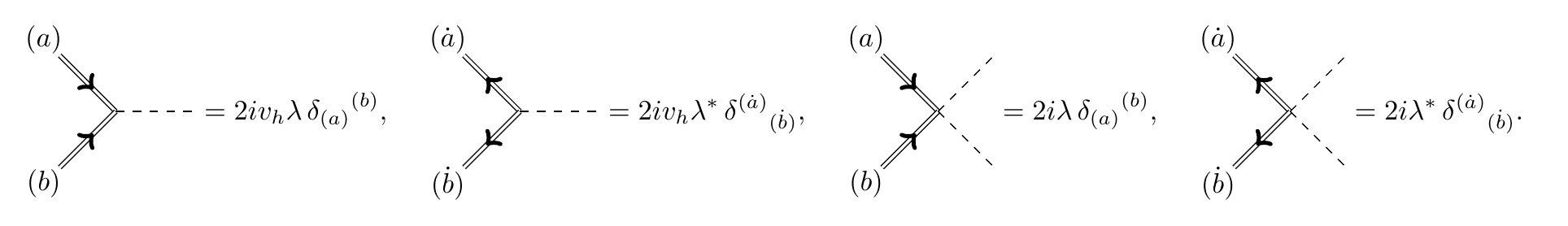}
    \caption{Feynman rules for $\psi$-Higgs interactions derived from the Higgs-portal term $\lambda \, \psi^{(a)} \psi_{(a)} (|\phi|^2 - \vev^2/2) + \hc$ in the interaction Hamiltonian for the $\R_j$ field. $\vev$ is the Higgs vev.}
    \label{tab:interactions}
\end{figure*}

In order to construct a model of general-spin DM, the DM particle must be stable. This can be achieved by imposing a $\mathbb{Z}_2$ symmetry. However, as will be shown later, for sufficiently high spin, an approximate $\mathbb{Z}_2$ symmetry appears. In the simplest scenario $\psi$ will be a color and electroweak singlet with vanishing hypercharge. The lowest dimensional operator in this case for any $\R_j$ field is the Higgs portal
\be\label{eq:portal}
    \Hcal_{\text{portal}}
    =
    -\lambda \, \psi^{(a)} \psi_{(a)} \parens{|\phi|^2 - \vev^2/2} + \hc,
\ee
where $\phi$ is the Higgs doublet and $\vev = \sqrt{2\angles{|\phi|^2}}$ is the Higgs vacuum expectation value (vev). The coupling constant $\lambda$ can generally be complex, with a real (imaginary) coupling corresponding to a parity-even (parity-odd) interaction. For $j = 0$, only the real part $\re{\lambda}$ contributes. The Feynman rules corresponding to~\eqref{eq:portal} are given in Fig.~\ref{tab:interactions}. 

The $\vev^2$ term is included in the portal interaction~\eqref{eq:portal} to avoid generating an extra contribution to the mass of $\psi$ in the phase where the electroweak symmetry is broken, i.e., $m$ will denote the pole mass of $\psi$. Moreover, for $j>1$, quadratic operators like $\psi^{(a)} \psi_{(a)} + \hc$ will introduce nontrivial momentum dependence into the denominator of the propagator (see Appendix~\ref{app:quadratic} for details), and should thus not be interpreted as simple mass terms. This is not surprising, because the formalism can at best provide an EFT description of general-spin particles. 
We define the \emph{effective cutoff scale} $\Lambdaunit$ as
\be\label{def:Lambda}
    \parens{\frac{m}{\Lambdaunit}}^{2j} \equiv \frac{|\lambda|}{4\pi},
\ee
which provides a rough upper limit for the validity of the theory. In particular, if $E$ is the typical energy of some process, we expect perturbative unitarity to be broken for that process when $E \simeq \Lambdaunit$. This is seen explicitly in Sec.~\ref{sec:pheno} for some specific processes. The portal operator \eqref{eq:portal} is thus effectively of dimension $4 + 2j$.

If $\psi$ would be charged under the SM gauge group, other types of $\psi$-SM interactions with the same dimension as the Higgs portal are allowed: interactions that couple two $\psi$ fields with SM gauge bosons have either operators of the schematic form $\psi^2 D^2$ or of the form $\psi^2 F$, where $D$ is the SM covariant derivative and $F$ is an SM field-strength tensor.\footnote{We remark that $\psi^2 F$ type interactions must couple states with different charges and are thus not allowed for SM-neutral fields even when $F$ is a $U(1)$ field strength. This is because $F_{ab}$ is symmetric in the spinor indices $a$, $b$, while $\psi_{a(c)}\psi_{b (d)}\epsilon^{(c)(d)}$ is always antisymmetric.} All other $\psi$-SM interactions are suppressed by further powers of $1/\Lambda$.

Consider now operators breaking the $\mathbb{Z}_2$ symmetry that may potentially render the general-spin fields unstable. In order for the general-spin particle corresponding to the field $\psi$ to decay into SM ones, $\Hcal_{\text{int}}$ must contain at least one term of the form
\be
    \Hcal_{\text{linear}}
    =
    \frac{1}{\Lambdalin^{\Delta_{\text{SM}} + j - 3}}
    \psi \Ocal_{\text{SM}},
\ee
where $\Ocal_{\text{SM}}$ is a local operator constructed out of SM fields only, $\Lambdalin$ is an energy scale of order $\Lambda$ and $\Delta_{\text{SM}}$ is the canonical dimension of $\Ocal_{\text{SM}}$. The SM fields are at most of spin 1 and all SM fermions carry a SM charge, i.e., neutral fermionic SM operators (for example, $\phi l$, with $l$ a lepton doublet) must at least involve a Higgs doublet. This implies that\footnote{These inequalities can be saturated, except for $j = 0$ for which $\Delta_{\text{SM}} \geq 2$, and for  $j = 1$ for which $\Delta_{\text{SM}} \geq 3$.}
\be
\Delta_{\text{SM}} \geq 
\left\{\begin{array}{ll}
    2j        & \mbox{ for bosons,}  \\
    2j + 3/2  & \mbox{ for fermions.}
\end{array}\right.
\ee
Thus, since the general-spin field can be treated as effectively $1+j$ dimensional, the decays will be induced by operators of dimension $1 + 3j$ for bosons and $5/2 + 3j$ for fermions. Comparing this with the effective dimension of the Higgs portal operator, $4 + 2j$, we find that the decays of $\psi$ are protected by an accidental $\mathbb{Z}_2$ symmetry for $j = 5/2$ and $j>3$, broken explicitly only at order $1/\Lambda^{\Delta_{SM} + j - 3}$. For lower spins, however, without the $\mathbb{Z}_2$ symmetry, the decays will not be suppressed with respect to the portal operator.

For completeness, let us briefly consider pointlike self-interactions of general-spin particles. Without the $\mathbb{Z}_2$ symmetry, the lowest order self-interactions are cubic $\psi^3$ for even spin. Several different $\psi^4$ operators can be constructed. For example, the operator 
$
    \psi^{(a)} \psi_{(a)} \,
    \psi^\dagger_{(b)} \psi^{\dagger\,(b)}
$
exists for any spin. We will assume here that self-interactions can be neglected for the DM phenomenology of $\psi$.

\section{Phenomenology of general-spin DM}
\label{sec:pheno}

General-spin particles, stabilized either by an exact or approximate $\mathbb{Z}_2$ symmetry, are potential candidates of DM. In the following we will focus on the simplest model in which such particles are SM singlets that interact with the visible sector via the Higgs portal~\eqref{eq:portal}. Despite the apparent simplicity of this interaction, it supports a relatively rich DM phenomenology which is partly owed to the complex phase of the coupling $\lambda$ allowing, e.g., for the suppression of direct detection signals.

\subsection{Dark matter abundance}

The evolution of the number density $n$ of the general-spin particle $\psi$ is described by the Boltzmann equation \cite{Gondolo:1990dk, Edsjo:1997bg}
\be\label{Boltzmann 1}
    \frac{\td n}{\td t} + 3H n
    = 
    -\angles{\sigma \vel} \parens{n^2 -n_{\rm eq}^2},
\ee
where $H$ is the Hubble rate and $\langle \sigma \vel\rangle$ is the thermally averaged annihilation cross section. The Hubble rate depends on temperature as $H=1.66 \sqrt{g_\ast(T)}T^2/\mPl$, where $g_\ast(T)$ is the effective number of relativistic degrees of freedom associated with the energy density. In the present model the general-spin  particle $\psi$ has the following annihilation channels: $\psi\psi\to f\bar{f}, W^+W^-, ZZ$ and $hh$, where $f$ is any charged SM fermion. The tree-level diagrams contributing to these processes are given in Fig.~\ref{fig:annihilation}.  Explicitly, the thermally averaged cross section is\footnote{We assume Boltzmann statistics and note that this can introduce $\mathcal{O}(1)$ errors with respect to Fermi or Bose statistics at high temperatures.}
\be
    \angles{\sigma \vel}
    =
    \frac{(2j+1)^2 T}{32\pi^4 n_{\text{eq}}^2}
    \int^\infty_{4m^2} \td s \, \sigma(s) \, (s-4m^2)\sqrt{s} \,
    K_1\left(\frac{\sqrt{s}}{T}\right).
\ee

The Boltzmann equation in Eq.~\eqref{Boltzmann 1} can be written in a more convenient form in terms of the yield, $Y = n/s$,  where $s$ is the entropy density $s(T)=(2\pi^2/45)g_{\ast s}(T)T^3$:
\be\label{Boltzmann 2}
    \frac{\td Y}{\td x} 
    =
    -\frac{\angles{\sigma \vel} s}{xH}
    \parens{Y^2-Y_{\rm eq}^2},
\ee
with $x = m/T$. The $g_{\ast s}(T)$ is the effective number of relativistic  degrees of freedom contributing to the entropy density. 

In the next two subsections we will study two different DM production mechanisms, the \emph{freeze-out} and \emph{freeze-in}.

\subsubsection{From freeze-out}

When the DM abundance is produced through the freeze-out mechanism, it is assumed that DM is initially in thermal equilibrium with the SM thermal bath. When the temperature drops bellow the mass of the DM, it becomes nonrelativistic and its equilibrium number density Boltzmann suppressed. Soon after, the number density will not be able to track its equilibrium value. Eventually, the DM annihilation rate will not be able to keep up with the expansion rate of the Universe and the DM freezes out. The last phase determines the DM abundance and can be approximately described by neglecting the exponentially small equilibrium yield $Y_{\rm eq}$ in the left-hand side of Eq.~\eqref{Boltzmann 2},
\be
     \frac{\td Y}{\td x}
     \simeq
     -\frac{\angles{\sigma \vel} s}{xH} Y^2,
\ee
which can be integrated to obtain the DM abundance,
\be
    \Omega h^2 = \frac{8.7\times 10^{-11}}{\si{GeV^2}} \left(
        \int^{\infty}_{x_f}\td x
        \langle \sigma \vel\rangle \frac{g_{\ast s}(T)}{x^{2}} 
    \right)^{-1},
\ee
where we used that $\Omega h^2 = 1.38\times 10^{8} Y_0 \,m/\si{GeV}$, with $Y_0$ the present yield, and $x_f = m/T_f$, with $T_f$ the freeze-out temperature, that is the temperature at which the DM number density starts to deviate from the equilibrium number density. We set it to $T_f = m/20$ in our analysis, thus neglecting its logarithmic dependence on the cross section. 

The magnitude of the interaction strength $\lambda$ required to produce the observed DM abundance, $\Omega_{\rm DM} h^2 = 0.120$~\cite{Aghanim:2018eyx}, for a given mass $m$ of the higher-spin particle is depicted in Fig.~\ref{fig:panel-plus-one} for a parity-even portal, and in Fig.~\ref{fig:panel-minus-one} for a parity-odd portal.

In order for the EFT approach to hold at the typical energy scales involved in our calculations, we require that $v_h < \Lambdaunit$ and $m < \Lambdaunit$, where the latter is equivalent to requiring perturbativity, $\lambda < 4 \pi$. These conditions will be translated into lower and upper bounds on the higher-spin particle mass. For $j\to\infty$, when demanding $\lambda \lesssim 4\pi$, the first condition translates into $m > v_h$, while lower spins allow lower masses. For low values of spin, the more stringent lower bounds on the mass come from direct detection or collider bounds, as will be discussed below.

In order to acquire the correct relic abundance the portal coupling has to be increased as the $m$ grows beyond $m_h$, as can be seen in Figs.~\ref{fig:panel-plus-one} and~\ref{fig:panel-minus-one}. Thus, as one keeps increasing $m$ one must violate perturbative unitarity at some point. We estimate the upper bound on $m$  using the following expressions for annihilation cross sections for different annihilation channels, in the limit $s \to 4 m^2$ and with $m \gg m_f, m_V, m_h$:
\begin{subequations}
\begin{align}
    \sigma_{\psi\psi \to \bar{f}f} \vel
    &\sim
    \frac{|\lambda|^2 m_f^2}{8\pi(2j + 1)  m^4} \\
    &\quad \times \Big[1 + (-1)^{2j} c_{2\theta} + \frac{2}{3}j(j+1) \vel^2\Big],
    \nonumber\\
    \sigma_{\psi\psi \to VV} \vel
    &\sim
    \frac{ \eta_V |\lambda|^2}{4\pi(2j + 1) m^2} \\
    &\quad \times \Big[1 + (-1)^{2j} c_{2\theta} + \frac{2}{3}j(j+1) \vel^2\Big],
    \nonumber\\
    \sigma_{\psi\psi \to hh} \vel
    &\sim
    \frac{|\lambda|^2}{8\pi(2j + 1) m^2} \\
        &\quad \times \Big[1 + (-1)^{2j} c_{2\theta} + \frac{2}{3}j(j+1) \vel^2\Big],
    \nonumber
\end{align}
\label{eq:non-rel-sigma}    
\end{subequations}
where $\eta_W = 2\eta_Z = 1$,  $c_{2\theta} \equiv \re{\lambda^2}/|\lambda|^2$ accounts for the complex phase of the coupling and, in the last equation, we have neglected terms proportional to $|\lambda|^4$. One acquires a different bound on the maximal $m$ depending whether the annihilation is an s wave ( $\sim v_{\rm rel}^0$) or p wave ($\sim v_{\rm rel}^2$).  
By imposing $\lambda \lesssim 4\pi$ and $\angles{\sigma \vel} \simeq \SI{3e-26}{cm^3 s^{-1}}$, which is the typical value required to produce the correct relic density, we obtain:
\be
    m \lesssim
    \begin{cases}
        \frac{1}{\sqrt{2j + 1}}\SI{100}{TeV} 
        &\text{if} \quad (-1)^{2j} c_{2\theta} > -1,
        \\
        \sqrt{\frac{j(j+1)}{2j + 1}} \SI{20}{TeV}
        &\text{if} \quad (-1)^{2j} c_{2\theta} \simeq -1.
    \end{cases}
\ee
This agrees with the known limit of order $\SI{100}{TeV}$ on the mass of thermally produced DM particles~\cite{Griest:1989wd}.

\subsubsection{From freeze-in}

In freeze-in production, the DM coupling to the SM thermal bath must be so weak that DM never thermalizes. Such feeble interactions are extremely difficult to detect experimentally (for a review see~\cite{Bernal:2017kxu}). The DM abundance, which is assumed to be negligible after inflation, is produced via scattering of SM particles into the dark sector. If the production cross section decreases with energy, typically as $s^{-1}$, the bath ceases to produce DM particles when the temperature of the thermal bath falls bellow the DM mass and most of the DM production will happen at this infrared (IR) limit near the DM mass. This is referred to as \emph{IR freeze-in}~\cite{Hall:2009bx} and, in this case, the DM yield is independent of physics in the UV, e.g., the reheating temperature. On the other hand, if the production cross section is not decreasing with temperature, most of the DM is produced in the UV regime~\cite{Hall:2009bx,Elahi:2014fsa}. Thus, in \emph{UV freeze-in} the DM yield is sensitive to UV physics and will depend on the highest temperatures in the early Universe.

The qualitative behavior of the freeze-in production depends on the high-energy behavior of the annihilation cross-section. The $s \to \infty$ asymptotic behavior of the annihilation channels (given in Appendix~\ref{app:xs-formulas})  is
\begin{subequations}
\begin{align}
    \sigma_{\psi\psi\to\bar{f}f}
    &\sim
    \frac{
        |\lambda|^2 m_f^2 (1 + \delta_{j0})
    }{
        (2j + 1)^2 \pi m^{4j}
    } s^{2j - 2},
    \\
    \si\sigma_{\psi\psi\to VV}
    &\sim
    \frac{
        \eta_V |\lambda|^2 (1 + \delta_{j0})
    }{
        2(2j + 1)^{2} \pi m^{4j}
    } s^{2j - 1},
    \\
    \sigma_{\psi\psi\to hh}
    &\sim
    \frac{
        |\lambda|^2 (1 + \delta_{j0})
    }{
        4 (2j + 1)^{2} \pi m^{4j}
    } s^{2j - 1} \nonumber\\
    &\qquad + \frac{2 |\lambda|^4 \vev^4 G_j}{(2j + 1)^2  \pi m^{8j}}
    s^{4j - 3},
\end{align}
\end{subequations}
where $\eta_W = 2\eta_Z = 1$ and
\be
    G_j = \frac{(2j - 2)! (2j)! + (-1)^{2j} [(2j - 1)!]^2}{(4j-1)!}.
\ee
The total annihilation cross section for $j=0$ behaves as $\sigma\sim s^{-1}$ in the UV. The freeze-in production is then dominated by the low-temperature regime where the DM is nonrelativistic. For spins $j>0$, the energy dependence is stronger, $\sigma\sim s^{n}$, with $n\geq 0$. Thus, for the Higgs portal~\eqref{eq:portal} considered here, the freeze-in takes place in the UV regime with scalars being the only exception.

Below, when estimating the DM abundance for UV freeze-in, we will assume that the visible sector will be instantaneously reheated to temperature $T_{\rm RH}$. However, this assumption can be violated if the universe is heated by inflaton decays, as a subdominant fraction of visible matter at temperatures higher than $T_{\rm RH}$ will be produced before the complete decay of the inflaton~\cite{McDonald:1989jd,Chung:1998rq,Giudice:2000ex,Chen:2017kvz,Garcia:2017tuj}. In this case, due to higher densities and UV-enhanced production, the bulk of the DM may be produced before reheating. Moreover, if the visible sector is not thermalized in the beginning of this epoch, the energies of the SM particles will be generally higher than in a thermal bath -- of the order of the inflatons mass -- further enhancing DM production~\cite{Harigaya:2013vwa,Garcia:2018wtq}. If the inflaton is nonminimally coupled, couplings to higher dimensional operators will be naturally generated enabling DM production through direct decays of the inflaton~\cite{Gorbunov:2010bn,Gorbunov:2012ij,Bernal:2020qyu}. For example, in Starobinsky inflation, DM production through direct inflaton decays may dominate over the freeze-in production already when $j>1$~\cite{Bernal:2020qyu}. The DM abundance in UV freeze-in may be further affected by the cosmological background~\cite{Bernal:2019mhf,Bernal:2020bfj}. In all, in order to avoid details related to UV physics, we adapt the instantaneous reheating approximation. However, it must be kept in mind that this approximation ignores DM production before the complete decay of the inflaton and may thus predict stronger couplings to produce the observed DM abundance.

In freeze-in, the DM number density is always much smaller than its thermal value. We can therefore ignore DM annihilation in the Boltzmann equation~\eqref{Boltzmann 2} which is now approximately
\be
    \frac{\td Y}{\td x}
    \simeq
    \frac{\angles{\sigma \vel} s}{xH} Y_{\rm eq}^2.
\ee
Analogously to the freeze-out case, this equation can be directly integrated, giving the DM abundance
\be
    \Omega h^2
    =
     3.4\times 10^{25} \, c_{j}^2 (2j+1)^2
    \int^{T_{\rm RH}}_{T_{\rm min}} \frac{\td T\, m \langle \sigma \vel\rangle }{\sqrt{g_\ast}g_{\ast s}},
\ee
where $c_{j}=1$ for bosons and $3/4$ for fermions, $T_{\rm RH}$ is the reheating temperature, and $T_{\rm min} \simeq 1.36 m$ is the temperature when the production becomes inefficient due to the Boltzmann suppression of the production rate~\cite{Gabrielli:2015hua}. For $j>0$, we set $T_{\rm min} = 0$. For the EFT approach to be valid, $T_{\rm RH} \ll \Lambdaunit$ must hold.

For $j\geq 1/2$, the present day abundance can be written as:
\begin{align}\label{UV freeze-in abundance}
    \Omega h^2
    =
    2.2\times 10^{21}
    \Bigg[\
        &A_j |\lambda|^2 \parens{\frac{\TRH}{m}}^{4j - 1}    
    \nonumber\\
   +\   &B_j |\lambda|^4 \frac{\vev^4}{m^4} \parens{\frac{\TRH}{m}}^{8j - 5}
    \ \Bigg],
\end{align}
where we took $g_{\ast s}(T_{\rm RH}) = g_{\ast s}(T_{\rm RH}) = 106.75$ and defined
\bea
    A_j &= \frac{2^{4j+2} \, (2j)! (2j+1)! }{4j - 1},
    \\
    B_j &= \frac{2^{8j-1} \, (4j-1)! (4j-2)! G_j}{8j - 5} .
\eea
For spins $1/2$ and $1$, the $|\lambda|^2$ term in Eq.~\eqref{UV freeze-in abundance} dominates. For $j\geq 3/2$, the $|\lambda|^4$ term in Eq.~\eqref{UV freeze-in abundance} can be relevant as its $T$ dependence becomes stronger than that of the $|\lambda|^2$ term. The $|\lambda|^4$ term is, however, suppressed by extra powers of the couplings. In particular, fixing $\Omega_{\rm DM} h^2 = 0.12$ we obtain that the second term is dominant for spin $j < 20$ only when $m T_{\rm RH}^3 \lesssim \si{GeV^4}$ and can thus be ignored for realistic models of reheating unless the spin is very high. Thus, as usual for freeze-in, the required coupling,
\be
    |\lambda| \simeq 7\times 10^{-12} \frac{1}{\sqrt{A_{j}}}\left(\frac{m}{T_{\rm RH}}\right)^{2j-1/2},
\ee
is too weak to be observable with current or planned experiments. As discussed earlier, accounting for DM production before the reheating is complete, will likely lower the required $|\lambda|$ even more.

\subsection{Collider constraints}

 The collider constraints of our effective framework are quite similar to usual Higgs-portal DM models \cite{Baglio:2015wcg, Craig:2014lda, Djouadi:2012zc, Djouadi:2011aa}. In this class of models  the only way of producing DM in colliders is by first producing Higgs bosons, either on shell or off shell, that subsequently dacays into DM: $pp\to h ~ X\to \psi\psi ~ X$, where $X$ represents visible SM states.  The prospects of a DM signal  then   crucially depends  on the mass of the DM. If the DM mass is $\leq m_h/2$, the Higgs boson can decay to DM on shell, which is an invisible decay. 
The SM  Higgs boson decays predominantly to visible channels. 
The only invisible decay channel of the Higgs boson is to the neutrinos, ${\rm BR}_{\rm SM}(h\to {\rm inv})={\rm BR}_{\rm SM}(h\to 4\nu)\simeq 10^{-3}$, and can be neglected.  
BSM contributions can significantly alter the invisible decay rate of the Higgs boson. 
If $\psi$ is heavier than $m_h/2$, the Higgs boson in  $pp\to h ~ X\to \psi\psi ~ X$ has to be virtual. The DM production process in this case is suppressed by $|\lambda |^2$ and the production rate will be small. 

\begin{figure*}
    \centering
    \includegraphics[width=0.8\linewidth]{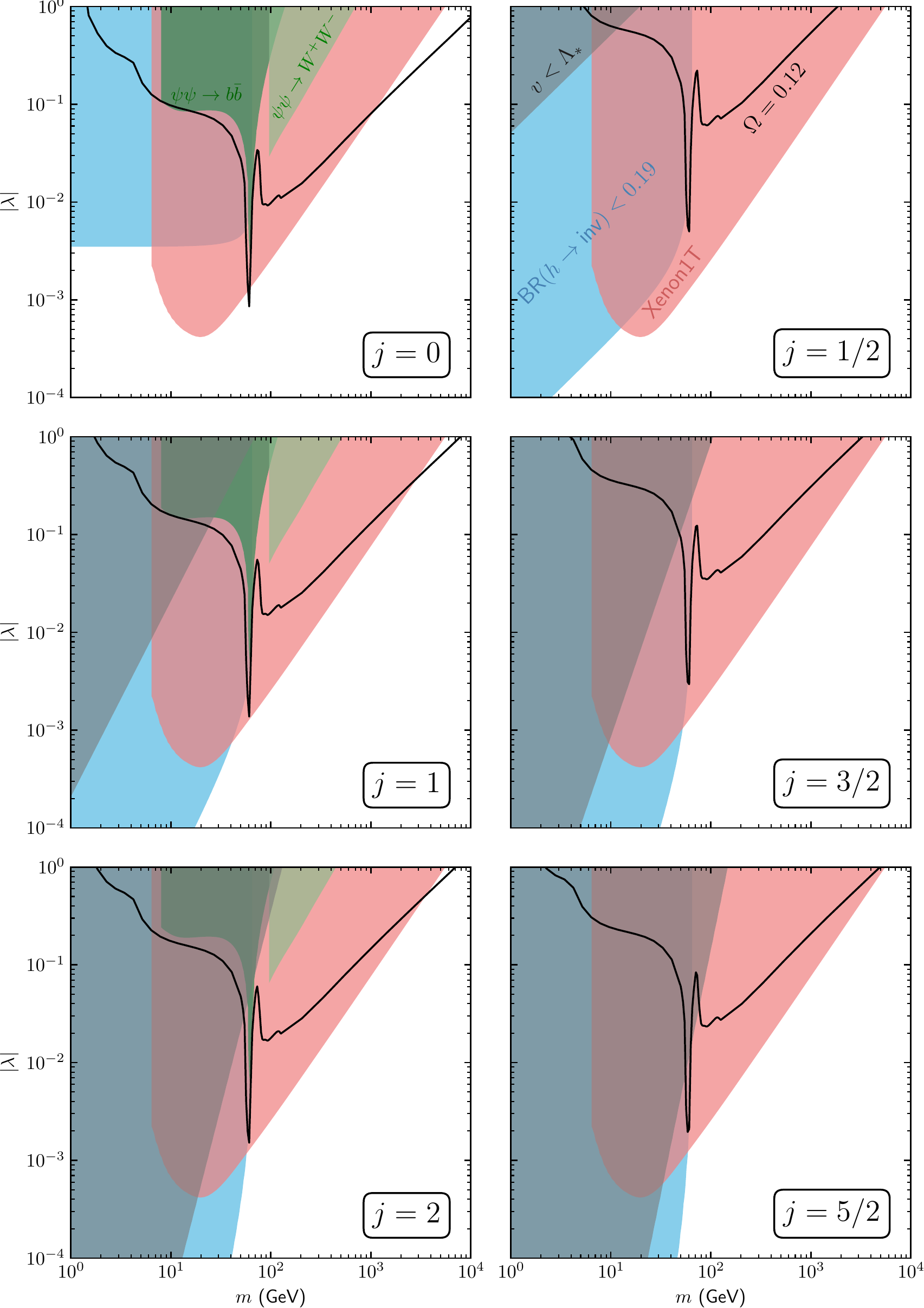}
    \caption{Excluded regions in $(m, |\lambda|)$ space from the CMS bounds on the branching ratio $\BR{h \to \text{inv}}$ (blue), from the Xenon1T limit on $\sigma_N$ (red), from the indirect limits on annihilation cross-sections from Ref.~\cite{Boddy:2019qak} (green), and from the perturbativity requirement $\vev < \Lambdaunit$ (gray),
    together with the lines of $(m, |\lambda|)$ values that give the correct relic abundance in a freeze-out scenario (black), for $c_{2\theta} = 1$ (corresponding to a vanishing complex phase $\theta = 0$ of $\lambda$) and different values of the spin $j$.}
    \label{fig:panel-plus-one}
\end{figure*}

\begin{figure*}
    \centering
    \includegraphics[width=0.8\linewidth]{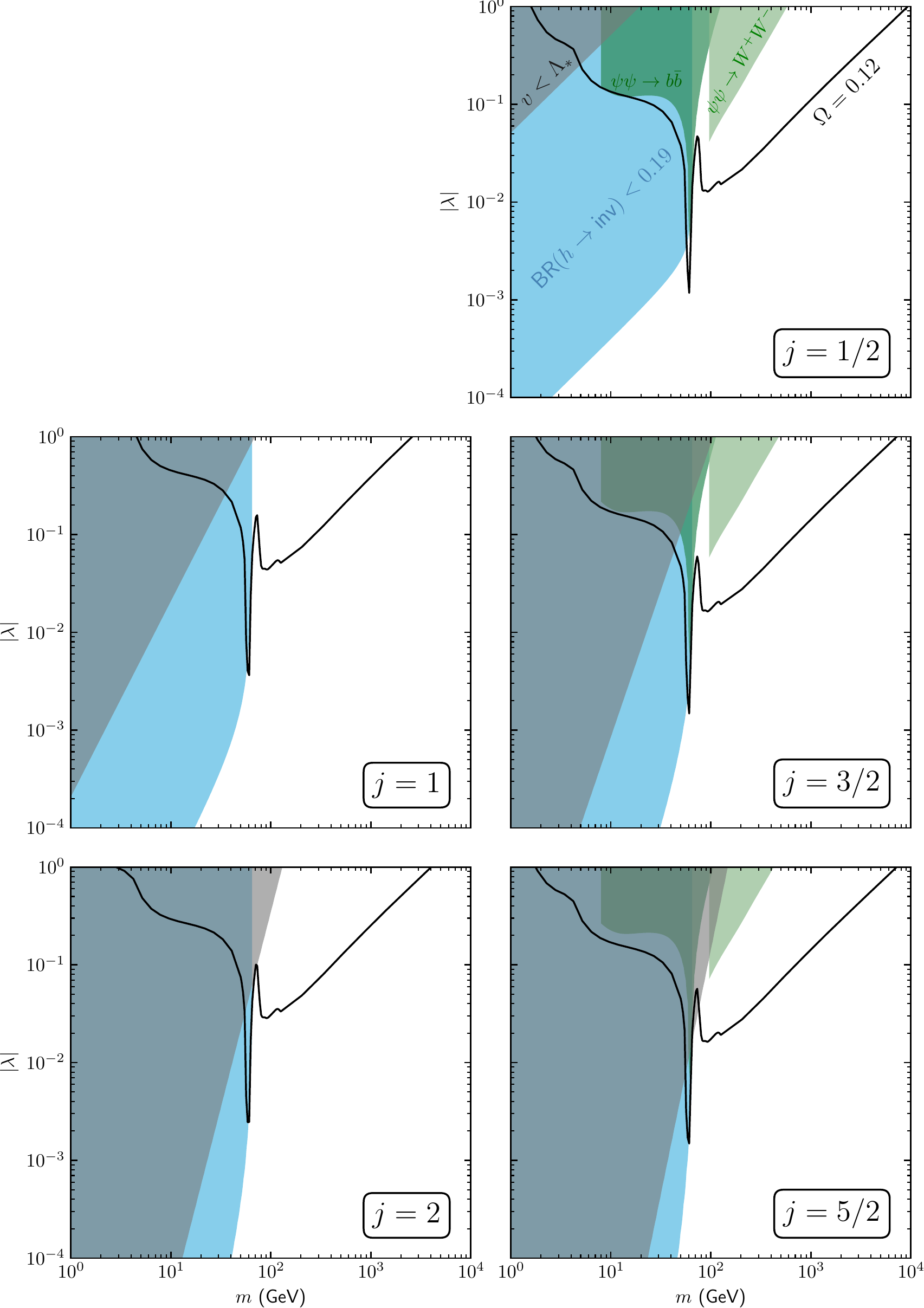}
    \caption{Excluded regions in $(m, |\lambda|)$ space from the CMS bounds on the branching ratio $\BR{h \to \text{inv}}$ (blue), from the indirect limits on annihilation cross-sections from Ref.~\cite{Boddy:2019qak} (green), and from the perturbativity requirement $\vev < \Lambdaunit$ (gray),
    together with the lines of $(m, |\lambda|)$ values that give the correct relic abundance in a freeze-out scenario (black), for $c_{2\theta} = -1$ (corresponding to complex phase of $\theta = \pi/2$ for $\lambda$) and for different values of the spin $j$.}
    \label{fig:panel-minus-one}
\end{figure*}

In the present model the invisible decay of the Higgs boson can be modified, if kinematically allowed, due to new decay channel $h\to\psi\psi$. The corresponding branching ratio is
\begin{equation}\label{brancing ratio to psi psi}
    {\rm BR}(h\to\psi\psi)=\frac{\Gamma(h\to\psi\psi)}{\Gamma_{h}+\Gamma(h\to \psi\psi)},
\end{equation}
with 
\bea
    \Gamma_{h \to \psi\psi}
    &=
    \frac{|\lambda|^2 \vev^2 m^2}{2^{2j+3} \pi m_h^3} 
    \\
        &\times\Bigg[\parens{\frac{m_h^2}{m^2} - 2+\frac{m_h^2}{m^2} \sqrt{1 - \frac{4 m^2}{m_h^2}}}^{2j+1}\\
        &- \parens{\frac{m_h^2}{m^2} - 2-\frac{m_h^2}{m^2} \sqrt{1 - \frac{4 m^2}{m_h^2}}}^{2j+1}
    \Bigg]\\
    &
     +\frac{|\lambda|^2\vev^2}{4\pi m_h}\sqrt{1 - \frac{4 m^2}{m_h^2}}(2j+1)(-1)^{2j} c_{2\theta},
\eea
where $m_h$ is the Higgs mass. The SM Higgs decay width to visible channels is $\Gamma_h=4.07$ MeV \cite{Dittmaier:2011ti}. 
The branching ratio for invisible decays has been determined experimentally to be less than $0.19$ at $95\%$ C.L.~\cite{Sirunyan:2018owy}. We impose this bound on the ${\rm BR}(h\to\psi\psi)$. This  excludes regions in the $(m, |\lambda|)$ plane for general spin $j$, as shown in Figs.~\ref{fig:panel-plus-one} and~\ref{fig:panel-minus-one}, for $c_{2\theta} = 1$ and $c_{2\theta} = -1$, respectively.

In order for perturbation theory not to be broken at the electroweak scale, it is conservatively required that $\vev$ is below the effective cutoff scale $\Lambdaunit$. Figures~\ref{fig:panel-plus-one} and~\ref{fig:panel-minus-one} shows the region excluded by this condition. 

In the case of parity-odd couplings, i.e., for purely imaginary $\lambda$, the invisible decay of the Higgs boson yields the strongest lower bound on $m$ for low spins. For $j = 1/2$ and $c_{2\theta} = -1$, one gets the lowest mass consistent with the freeze-out scenario to be $m \simeq m_h/2$. For spins higher than $j=5/2$, the Higgs boson decay takes place close to the nonperturbative regime, and, thus, the DM mass must be larger for the EFT to be trusted.

\subsection{Direct detection}

The direct detection prospects depend on the production mechanism. The portal coupling for frozen-in DM are typically too weak to be constrained by direct detection. However, the portal coupling needed for production via freeze-out may be sufficiently large to be detected in direct detection experiments. In the simplest weakly interacting massive particles (WIMP) models both the DM annihilation and direct detection cross sections depend on the same coupling, and direct detection excludes DM masses around the electroweak scale~\cite{Arcadi:2019lka}. However, there exist models where WIMP direct detection cross section is suppressed. These include secluded DM models where the DM annihilation and the direct detection cross sections depend on different parameters and the stringent constraints can be avoided~\cite{Pospelov:2007mp, Arcadi:2016qoz}. Direct detection bounds can be avoided also in models where DM co-annihilations~\cite{Casas:2017jjg} or some cancellations in direct detection amplitude are utilized~\cite{Arcadi:2016kmk}. Most importantly, the WIMP direct detection cross section can be suppressed in models where the DM is a pseudo-Goldstone boson~\cite{Gross:2017dan, Huitu:2018gbc, Alanne:2018zjm, Azevedo:2018oxv, Kannike:2019wsn, Kannike:2019mzk,  Cline:2019okt, Alanne:2020jwx}. In these models the tree-level direct detection amplitude vanishes at zero momentum transfer limit due to derivative couplings of the pseudo-Goldstone DM. Intriguingly, also our effective framework allows for the cancellation of the tree-level direct detection amplitude.

The general-spin particle $\psi$ couples to nucleon through Higgs bosons. The $\psi$ is  therefore potentially subject to stringent direct detection bounds. The current bounds are from XENON1T~\cite{Aprile:2018dbl}. The SM Higgs boson $h$ couples to nucleon $N$ through the following effective coupling:
\be
    \mathcal{L}_{hN} =\frac{f_N m_N}{v} \overline{N} N h,
\ee
where $m_N=0.946$ GeV is the nucleon mass and $f_N = 0.3$ is the effective nucleon coupling~\cite{Alarcon:2011zs, Alarcon:2012nr, Cline:2013gha}.

In the limit of zero momentum transfer, the tree-level DM-nucleon scattering cross section is
\be
    \sigma_N = 
    \frac{
        2 m_N^2 \mu_N^2 f_N^2 |\lambda|^2
    }{
        \pi m_h^4 m^2
    }
    \Big[
        1 + c_{2\theta}
        + \frac{4}{3}j(j+1)\frac{\mu_N^2 \vel^2}{m^2}
    \Big],
    \label{eq:sigma-N}
\ee
where $\mu_N \equiv m m_N / (m + m_N)$ denotes the DM-nucleon reduced mass. 

In Fig.~\ref{fig:panel-plus-one} we show the excluded region in $(m, |\lambda|)$ space induced by the $m$-dependent 90\% C.L. limit on $\sigma_N$ measured by the XENON1T Collaboration~\cite{Aprile:2018dbl}, for real coupling ($c_{2\theta} = 1$). In this case, freeze-out DM with masses below $\SI{1}{TeV}$ are excluded for any spin. This bound can be saturated only when $j = 0$, while fermions with $j < 11$ are completely excluded by it in conjunction with the perturbativity condition $|\lambda| < 4\pi$.

The cross section~\eqref{eq:sigma-N} reveals an important feature: the direct detection cross section vanishes for the purely imaginary (parity-odd) portal coupling and zero velocity.\footnote{However, DM particles passing through a detector at the Earth have non vanishing velocities. Although the local velocity distribution of DM particles is highly uncertain (e.g. \cite{Bozorgnia:2016ogo, Benito:2016kyp, Herzog-Arbeitman:2017fte, OHare:2019qxc}), assuming a velocity $v_{\rm rel}=\SI{220}{km/s}$ (which corresponds to the Sun's circular velocity) gives $\sigma_N < \SI{e-47}{cm^{-2}}$ for $\SI{6}{GeV} < m < \SI{1}{TeV}$, well below current experimental limits.} Thus, setting $c_{2\theta} \simeq -1$ allows one to escape the stringent direct detection bounds, and they become irrelevant for  Fig.~\ref{fig:panel-minus-one}, in which we have set $c_{2\theta} \simeq -1$.

\subsection{Indirect detection}

The relevant indirect detection constraints arise due to DM annihilations in spheroidal dwarf galaxies orbiting the Milky Way. The DM in these structures is cold and it is therefore justified to take the vanishing momentum limit in annihilation cross sections. The DM annihilations in dwarf galaxies produce a gamma-ray flux that depends on the density profile of the DM halo. This effect is described by the J factor. We use the constraints on annihilation channels $\psi\psi\to b\bar{b}$ and $\psi\psi\to W^+W^-$ that are based on J-factors given in Ref.~\cite{Boddy:2019qak}.

The annihilation cross sections of relevant channels at vanishing momentum are given in Eq.~\eqref{eq:non-rel-sigma}, with particle masses omitted for simplicity. Full expressions were used in the numerical analysis. The 95\% C.L. limits on the DM annihilation cross section into $b\bar{b}$ and $W^+W^-$~\cite{Boddy:2019qak} are shown in Figs.~\ref{fig:panel-plus-one} and~\ref{fig:panel-minus-one}. As above, two qualitative extremes can be distinguished, depending on whether the portal coupling is real or imaginary.  The real portal coupling corresponds to s-wave ($\sim v_{\rm rel}^0$) DM annihilation for bosons and p wave annihilation ($\sim v_{\rm rel}^2$) for fermions. As the velocity suppressed bounds are extremely weak, the indirect detection constraints for fermions do not appear in the right panels of Fig.~\ref{fig:panel-plus-one}. For the imaginary portal coupling the situation is reversed. Now the bosonic DM annihilations are p wave and the fermionic DM is s wave. The indirect detection constraints for bosonic DM is now velocity suppressed and does not show in Fig.~\ref{fig:panel-minus-one}. In all cases considered, the thermal DM abundance is not constrained by indirect detection.

\section{Results and discussion}
\label{sec:results}

We now briefly describe our results for each individual spin and discuss the relation with previous works, whenever they exist. As a general feature, the necessary values of  coupling $|\lambda|$ for the correct DM abundance to be generated through freeze-out are excluded by the bounds on invisible decays of the Higgs boson for $m < m_h/2$ and by direct detection experiments for $\SI{6}{GeV} < m < \SI{1}{TeV}$, unless $c_{2\theta} \simeq -1$. However, if $\lambda$ is near the imaginary axis, the DM-nucleon cross section is suppressed, and direct detection limits can be evaded, allowing for freeze-out DM with masses $m > m_h/2$. On the other hand, no experimental bounds apply to the values of $\lambda$ required for the freeze-in mechanism to work, except for very high spin $j$.

\begin{description}[leftmargin=30pt, style=sameline]
\item[$j=0$] Among the bosons, the scalar is the one with the mildest bounds on $|\lambda|$ from direct and indirect detection as well as from the invisible decays of the Higgs boson, as compared with cases with other spins in which $c_{2\theta}$ is away from $-1$. However, since $\lambda$ is real for a scalar particle, this is the only spin for which the $c_{2\theta} \simeq -1$ mechanism for the suppression of the direct detection cross section does not exist. Another unique feature of spin 0 is that the bound from the Higgs decays stays almost constant (at $\lambda_{\text{max}} \simeq \num{4e-2}$) for $m < m_h/2$.

Real and complex scalar singlet DM has been studied in Refs.~\cite{Silveira:1985rk, Burgess:2000yq, Kanemura:2010sh, Cline:2013gha, Arcadi:2019lka} and in Refs.~\cite{McDonald:1993ex, Barger:2008jx, Gabrielli:2013hma}, respectively. Our spin-0 field $\psi$ is just the conventional real scalar field, with portal coupling
\be
    \Lcal_{\text{portal}} = 2 \lambda |\phi|^2 \psi^2.
\ee
Our analytic results for the Higgs invisible width and DM-nucleon cross section agree with those given in Ref.~\cite{Kanemura:2010sh}.

\item[$j=\frac{1}{2}$] This is the lowest spin for which the following three features appear: the $c_{2\theta} \simeq -1$ suppression mechanism for the DM-nucleon cross section exists; the limits from the Higgs invisible decays become stronger for lower masses; and perturbative unitarity is broken at some finite energy. These three features are present for any spin higher than 1/2.

Spin-1/2 DM has been considered in Refs.~\cite{Kanemura:2010sh, Baek:2011aa, LopezHonorez:2012kv, Fedderke:2014wda, Balaji:2018qyo, Arcadi:2019lka}. Our spin-1/2 particle is a Majorana fermion. In terms of the usual four-component spinor $\Psi$ representing such a fermion, the portal interaction becomes
\be
    \Lcal_{\text{portal}}
    =
    \frac{1}{m} |\phi|^2 
    \overline\Psi \parens{\re{\lambda} - i \im{\lambda} \gamma^5} \Psi,
    \label{eq:majorana}
\ee
As in the scalar case, our analytic results for the Higgs invisible width and DM-nucleon cross section agree with those given in Ref.~\cite{Kanemura:2010sh} for real $\lambda$. Two of the properties of this model become apparent in view of Eq.~\eqref{eq:majorana}: the imaginary part of $\lambda$ is associated with pseudoscalar DM-Higgs interactions, which are known to generate a small contribution to DM-nucleon scattering (see Refs.~\cite{LopezHonorez:2012kv, Fedderke:2014wda}); and perturbative unitarity is broken because the DM-Higgs interactions are generated by a dimension-5 operator, with a coefficient of order $\lambda/m$.

\item[$j = 1$] Our results here are similar to those for spin 1/2, but now our formulation does not coincide with the usual one. Spin-1 DM has been studied in Refs.~\cite{Hambye:2008bq, Kanemura:2010sh, Lebedev:2011iq, Gross:2015cwa, Karam:2015jta, Arcadi:2019lka} using a $(1/2, 1/2)$ field and in Ref.~\cite{Hernandez-Arellano:2018sen} using a $(1, 0) \oplus (0, 1)$ field. A direct analogy between our field and the $(1/2, 1/2)$ is harder to make. However, the physics should be the same in any case. For example, in Ref.~\cite{Arcadi:2019lka}, it is shown that spin-1 CP-preserving DM with $m_h/2 \neq m < \SI{1}{TeV}$ is excluded when its abundance is to be set by freeze-out. This also happens in our formulation, as can be seen in Fig.~\ref{fig:panel-plus-one}.

\item[$j = \frac{3}{2}$] We find here similar results as for the lower spins. Just as for spin-1, our formulation differs from the usual one.  In nonsupersymmetric context, spin-3/2 DM has been considered in~\cite{Kamenik:2011vy,Yu:2011by,Ding:2012sm,Ding:2013nvx,Dutta:2015ega,Khojali:2016pvu,Chang:2017dvm,Khojali:2017tuv,Goyal:2019vsw,Garcia:2020hyo}. These works use the Rarita-Schwinger formulation, in which the field irrep is $(1, 1/2) \oplus (1/2, 1)$, while the irrep we use is $\R_{3/2}$.\footnote{We have nevertheless listed all the effective operators that would be allowed for a $(1, 1/2) \oplus (1/2, 1)$ field in different SM gauge group irreps in Appendix~\ref{app:spin-32}.} Locally supersymmetric theories (i.e. supergravity) generally predict the existence of spin-3/2 particle called the gravitino, the supersymmetric partner of spin-2 graviton \cite{Freedman:1976xh, Deser:1976eh, Freedman:1976py}. In supergravity the gravitino is described as a Rarita-Schwinger field.  When the local supersymmetry is exact, the gravitino is massless.  The supersymmetry must be broken at low energies, and therefore, the gravitino acquires a mass through super-Higgs mechanism \cite{Cremmer:1982en}.  The gravitino can be stable or unstable, depending on whether it is the lightest supersymmetric particle (LSP) or not. If the gravitino is the LSP, it is stable and a possible DM candidate.  In fact, the gravitino was the first supersymmetric DM candidate proposed \cite{Pagels:1981ke, Khlopov:1984pf}. However, if the gravition thermalizes, the universe is overclosed if $m_{3/2}\gtrsim {\rm keV}$, which is in strong tension  with large-scale structure formation \cite{Kunz:2016yqy} and the Tremaine-Gunn limit \cite{Tremaine:1979we}.  Alternative mechanisms for the generation of  gravitino abundance exist where the gravitino does not thermalize, thus avoiding the above problems. One of these mechanisms produces gravitinos through thermal scatterings  after the inflation~\cite{Ellis:1984eq, Moroi:1993mb, Ellis:1995mr,Giudice:1999am, Bolz:2000fu, Cyburt:2002uv, Steffen:2006hw, Pradler:2006qh, Rychkov:2007uq, Kawasaki:2008qe, Ellis:2015jpg,Benakli:2017whb}. Another mechanism is to produce  gravitinos  through decays of other supersymmetric particles in a thermal bath, that is through freeze-in~\cite{Cheung:2011nn}. 

The spin-3/2 phenomenology of the effective approach we have adopted differs greatly from that of gravitino of the supergravity. In supergravity, the couplings of the gravitino are completely determined by the other couplings of the theory, such as gauge couplings. In contrast, the portal coupling of our effective approach is free and not related to other couplings of the model. The  spin-3/2 particle of the effective approach can safely thermalize and its abundance be produced through the usual freeze-out, unlike in case of the gravitino.  

\item[$j = 2$] Spin-2 DM has been studied in Refs.~\cite{Aoki:2016zgp,Babichev:2016hir,Babichev:2016bxi,Marzola:2017lbt} in the context of ghost-free bimetric gravity~\cite{Hassan:2011zd}.  In this case the field is a symmetric rank-2 tensor whose representation under the Lorentz group decomposes as $(1, 1) \oplus (0, 0)$. The ghost-free bimetric model is the only known realization of such a spin-2 field that does not contain a dynamical scalar ghost. It predicts gravitylike universal interactions between the heavy spin-2 field and other matter whose strength is controlled by a single coupling constant. Our irrep for spin-2, on the other hand, corresponds to a rank-4 tensor with the symmetries of the Weyl tensor, which transforms  as $(2, 0) \oplus (0, 2)$. 
\item[$j > 2$] The situation for even higher spins is similar to the one for $j \leq 2$. The experimental limits slowly become stronger as $j$ increases, since the relevant cross sections and decay widths grow with $j$. Spin-3 DM has been considered in Ref.~\cite{Asorey:2010zz}.
\end{description}

\section{Conclusions}
\label{sec:conclusions}

We have presented an EFT description of massive particles of any spin. In order to do this, we have used fields in the $(j, 0)$ irrep of the Lorentz group. Contrary to other options for describing higher-spin particles, these fields do not contain any unphysical degrees of freedom. The proposed EFT framework is, therefore, free of consistency problems, such as causality violation, except for the breaking of perturbative unitarity at high-enough energies above the cut off scale $\Lambda$.

A Lagrangian formulation for such fields is likely not to exist, at least without involving a complicated system of extra fields and constraints. This motivated us to use Weinbergs prescription~\cite{Weinberg:1964cn}, which directly produces the Feynman rules needed to compute any amplitude perturbatively. We have reformulated this theory using our own symmetric multispinor notation, which leads to a considerable simplification of the Feynman rules, rendering them easy to use for practical calculations.

As an application, we have used this framework to study DM of any spin. The minimal DM models, involving a SM singlet, already contain interesting phenomenology. First, if the particles spin is high enough, an accidental stabilizing $\mathbb{Z}_2$ symmetry arises, rendering it suitable to be a DM candidate. If the spin is low, this symmetry has to be imposed by hand. In both cases, the most relevant operator for phenomenology is the Higgs-portal coupling. Our model then depends on four (one discrete and three real) free parameters: the spin $j$, the mass $m$, the coupling constant modulus $|\lambda|$ and the phase $\theta$ of the coupling constant, with a purely real (imaginary) coupling corresponding to a parity-even (odd) portal.

We found that, for general $\theta$ and a sufficiently high mass, general-spin DM whose abundance is set through the freeze-out mechanism is allowed by the current experimental bounds, except for low-spin fermions. An intriguing feature arises for the parity-odd portal, that is, for $\theta = \pm \pi/2$. Then, direct detection bounds are avoided and the DM mass can be as low as $\SI{51}{GeV}$ for lower spins, while for higher spins the masses must be somewhat higher, mostly due to perturbative unitarity considerations.

For $j > 0$, we find that the freeze-in takes place in the UV, so that most of the DM is produced near the highest temperatures in the early universe. This is because cross sections grow as $E^{4j - 2}$ (or $E^{8j - 6}$ when $|\lambda|$ is large enough) with the center-of-mass energy $E$ of the process.

The general conclusion of our work is that the proposed framework represents a tool to address and to compute phenomenology of generic fields with higher spin. Our formalism is free of inconsistencies and allows us to use the EFT language to compute high- and low-energy observables involving particles with any spin. UV completion of the proposed framework remains, however, a mystery.

\vspace{5mm}
\noindent \textbf{Acknowledgement.} We thank Abdelhak Djouadi, Adam Falkowski, Camila Machado and Alessandro Strumia for discussions. 
This work was supported by the Estonian Research Council Grants No. MOBTTP135, No. PRG803, No. MOBTT5, No. MOBJD323 and No. MOBTT86, and by the EU through the European Regional Development Fund CoE program TK133 ``The Dark Side of the Universe." J.C.C. is supported by the STFC under Grant No. ST/P001246/1.

\appendix

\section{Symmetric multispinor index notation}
\label{app:notation}

In this appendix we define a general notation for the components of objects in any irrep of the Lorentz group, together with some convenient definitions for dealing with such objects. The basis of our notation lies in the well-known two-component spinor formalism, reviewed for example in Ref.~\cite{Dreiner:2008tw}. We briefly summarize its main ingredients here.

We denote indices for the $(1/2, 0)$ irrep with lowercase letters from the beginning of the latin alphabet: $a, b, \ldots$. Indices $(0, 1/2)$ irrep are denoted with the same kind of letters decorated with a dot: $\dot{a}, \dot{b}, \ldots$. Any of these indices can appear in the up or down positions, so the most general object $t$ one can write indexed by them is of the form
\be
    t^{a_1 \ldots a_k \, \dot{a}_1 \ldots \dot{a}_l}_{b_1 \ldots b_k \, \dot{b}_m \ldots \dot{b}_n}.
\ee
Such an object is called a \emph{multispinor}. A Lorentz transformation acts naturally on such an object with a $(1/2, 0)$ or $(0, 1/2)$ representation for each indices. Kronecker deltas $\delta^b_a$ or $\delta^{\dot{b}}_{\dot{a}}$ for two indices of the same type are covariant when the two indices are at different heights. The only covariant objects with two indices of the same type at the same height are the epsilon symbols $\epsilon_{ab}$, $\epsilon_{\dot{a}\dot{b}}$, $\epsilon^{ab}$, and $\epsilon^{\dot{a}\dot{b}}$, which are antisymmetric $2\times2$ matrices satisfying $\epsilon_{12} = -\epsilon^{12} = 1$. They are used to raise and lower indices.\footnote{The only exception to this rule is the $\epsilon$ symbol itself, for which $\epsilon^{ab} = -\epsilon^{ac}\epsilon^{bd}\epsilon_{cd}$.} 

Two indices of a different type at the same height can be converted into one vector index $\mu$ using the tensor $\sigma^\mu_{a\dot{a}}$, defined as $\sigma^0$ being the identity matrix and $\sigma^i$ for $i=1, 2, 3$ the Pauli matrices
\be
    \sigma^\mu_{a\dot{a}} = (1,\sigma^i), \quad
    \bar\sigma^\mu{}^{\dot{a}a} = \epsilon^{a b} \epsilon^{\dot{a} \dot{b} }\sigma^\mu_{b\dot{b}}  = (1,-\sigma^i).
\ee
They satisfy the identity
\be\label{eq:ssbar}
    (\sigma^{\mu}\bar{\sigma}^{\nu} + \sigma^{\nu}\bar{\sigma}^{\mu})^{a}_{b} 
=   2\eta^{\mu\nu} \delta_{a}^{b}.
\ee
In particular, we will denote vectors $p_\mu$ and derivatives in their two-component spinor form:\footnote{Note that there is a sign difference between our definition Eq.~\eqref{eq:vector-definition} and that of Ref.~\cite{Dreiner:2008tw}. This is because our definition gives simpler expressions in the simplified notation for general spin $j$. Furthermore, we will not make an independent definition for $p^{\dot{a}a}$, as in Ref.~\cite{Dreiner:2008tw}. Instead, in our case, $p^{a\dot{a}}$ is obtained by raising the indices of $p_{a\dot{a}}$ with $\epsilon$ symbols. This is equivalent to using Ref.~\cite{Dreiner:2008tw}'s definition of $p^{\dot{a}a}$ and then exchanging the places of the dotted and the undotted index.}
\be
    p_{a\dot{a}} = \sigma^\mu_{a\dot{a}} p_\mu,
    \quad
    \partial_{a\dot{a}} = \sigma^\mu_{a\dot{a}} \partial_\mu .
    \label{eq:vector-definition}
\ee
One commonly used convention for two-component spinor indices is that contractions of undotted indices should be made from in descending order, whereas for dotted ones they should be in ascending order, e.g., $u^{a} = \epsilon^{ab} u_{b}$. This has the advantage that expressions stay unambiguous when indices are suppressed. On the other hand, there are some expressions (usually involving traces) in which it is impossible to make all the indices explicit without violating this convention. In this paper, we choose to make the indices explicit, and thus, we only follow the convention when possible.

Components of fields in irreps of the Lorentz group are easily denoted in terms of the undotted and dotted indices. First, it should be noticed that any spin-$j$ irrep of $SU(2)$ can be viewed as a symmetric tensor product of the fundamental representation. The extension to the Lorentz group is straightforward: a $(l, r)$ irrep can be interpreted as a symmetric tensor product of $l$ $(1/2, 0)$ irreps times a symmetric tensor product of $r$ $(0, 1/2)$ irreps. In terms of indices, this corresponds to a field of the form,
\be
    \psi_{a_1 \ldots a_{2l} \, \dot{a}_1 \ldots \dot{a}_{2r}},
    \label{eq:field-indices}
\ee
which is totally symmetric in the undotted and in the dotted indices.

We will often encounter expressions with symmetrized tuples of indices $(a_1 \ldots a_{2l})$ or $(\dot{a}_1 \ldots \dot{a}_{2r})$. A convenient notation when the number of indices in such tuples is known from the context is to denote them by symmetric multi-indices $(a)$ or $(\dot{a})$. Since the indices contained in $(a)$ or $(\dot{a})$ are symmetrized, we call them \emph{symmetric multispinor indices}. In terms of symmetric multispinor indices, the fields of the form in Eq.~\eqref{eq:field-indices} are written as
\be
    \psi_{(a)(\dot{a})}.
\ee
All the indices of a multispinor $t$ can be converted into symmetric multispinor indices with $l = r = j$ by taking the product of $j$ copies of $t$ and symmetrizing indices. As an example, one can generate the multispinor $t^{(a)}_{(\dot{a})}$ from the multispinor $t^{a}_{\dot{a}}$ as
\be
    t^{(a)}_{(\dot{a})}
    =
    t^{(a_1}_{(\dot{a}_1} \ldots t^{a_n)}_{\dot{a}_n)}.
\ee
Applying this procedure to the $\epsilon_{ab}$ and $\epsilon^{ab}$ symbols gives rise to the generalized $\epsilon$ symbols $\epsilon_{(a)(b)}$ and $\epsilon^{(a)(b)}$, which can be used to raise and lower symmetric multispinor indices.

Some useful algebraic relations in the notation defined here are
\begin{gather}
    \epsilon_{(a)(b)} = (-1)^{2j} \epsilon_{(b)(a)}, \\
    \epsilon^{(a)(c)} \epsilon_{(c)(b)}  = \delta^{(a)}_{(b)}, \\
  \delta_{(a)}^{(a)} = 2j + 1, \\
    x^{(a)} y_{(a)} = (-1)^{2j} y^{(a)} x_{(a)}, \\
     p_{(a)(\dot{a})} p^{(\dot{a})(b)} = (p^2)^{2j} \delta_{(a)}^{(b)}, \label{eq:mult_p2}
\end{gather}
for bosonic $x$, $y$, and $p$.

Identities for traces and contractions of symmetric multispinor objects follow from the two basic properties:
\begin{itemize}
    \item  The multiplicative structure of rank-2 two-spinor tensors is preserved when they are promoted to the corresponding symmetric multispinor object, e.g.
\be\label{eq:prod_sym} 
    X_{(a)(\dot c)} Y^{(\dot c)(b)} = (XY)_{(a)}{}^{(b)}, 
\ee
where $(XY)_{a}{}^{b} = X_{a\dot c} Y^{\dot c b}$. The identity \eqref{eq:mult_p2} is a simple example of this rule. Note that additive structure is not preserved and thus identities like \eqref{eq:ssbar} do not easily generalize to multispinors.
    \item The trace of a rank 2 symmetric multispinor tensor is a complete homogeneous symmetric polynomial of the eigenvalues of the corresponding two-spinor tensor:
\be\label{eq:trace_sym}
    X_{(a)}{}^{(a)} = \frac{x_{+}^{2j+1} - x_{-}^{2j+1}}{x_{+} - x_{-}}, 
\ee
where $x_{\pm}$ are the eigenvalues of $X_{a}{}^{b}$ determined from
\be
    x^2 - X_{a}{}^{a} x + \det(X_{a}{}^{b}) = 0.
\ee
Eq.~\eqref{eq:trace_sym} can be proved by noting that the multispinor trace is basis independent (as expected) and then studying it in the basis where $X_{a}{}^{b}$ is diagonal.
\end{itemize}

For example, multispinor traces of products of momenta, such as $p_{(a)(\dot{a})} q^{(\dot{a})(a)}$, can be computed using the trace identities in the two-spinor formalism. The corresponding two-spinor object to be used in Eq.~\eqref{eq:trace_sym} in the current example is $X_{a}{}^{b} = p_{a \dot{a}} q^{\dot{a}b}$. The determinant of a vector is simply $|p_{a \dot{a}}| = p^2$ and the trace reads
$p_{a \dot{a}} q^{\dot{a}a} = 2 p\cdot q$. Applying Eq.~\eqref{eq:trace_sym} then gives that
\bea\label{eq:tr_pq}
    p_{(a)(\dot{a})} q^{(\dot{a})(a)} &= 
    \frac{\left(p\cdot q + \sqrt{(p\cdot q)^2 - p^2 q^2}\right)^{2j+1}}{2\sqrt{(p\cdot q)^2 - p^2 q^2}} \\
&\phantom{=}-  \frac{\left(p\cdot q - \sqrt{(p\cdot q)^2 - p^2 q^2}\right)^{2j+1}}{2\sqrt{(p\cdot q)^2 - p^2 q^2}} .
\eea
Of course, any multispinor trace has an equivalent expression as a homogeneous polynomial of scalar products of the involved momenta, although these polynomials can get rather complicated when more than 2 momenta are involved. Carrying on with the example, \eqref{eq:tr_pq} can be recast as
\bea
&    p_{(a)(\dot{a})} q^{(\dot{a})(a)} 
   = \sum_{k = 0}^{\lfloor j \rfloor} a_k  (p \cdot q)^{2j - 2k} (-p^2 q^2)^k, 
\eea   
with
\bea
    a_k =
    \sum_{n = k}^{\lfloor j \rfloor} 
        \begin{pmatrix}2j + 1 \\ 2n + 1\end{pmatrix}
        \begin{pmatrix}n \\ k\end{pmatrix}.
\eea

\section{Quantization of second-order fermions}
\label{app:quantization-fermions}

A possible alternative to the usual Dirac and Rarita-Schwinger formulation based on single derivative Lagrangians is to use a two derivative formulation analogous to the bosonic case. In this appendix, we comment on the pitfalls of such an approach. One can try to write down a second-order theory by constructing a second-order free Lagrangian, since the lowest number of derivatives allowed by Lorentz invariance for the kinetic term of a $\R_j$ field with $j > 1/2$ is 2. For example, the simplest Lagrangian of this kind is
\be
    \Lfree^{\text{second-order}}
    =
    -\frac{1}{2} \psi^{(a)} (\square + m^2) \psi_{(a)} + \hc.
\ee
Apart from the problems outlined in Sec.~\ref{sec:free} the theory defined in this way will lead to inconsistencies in the Hilbert-space representation of $\psi$. Let us use a simplified notation in which all indices of fields, including the spacetime point, are collected in one multi-index, denoted by a greek letter $\alpha$, $\beta$, etc. Canonical quantization fixes the (anti)commutation relations between $\psi_\alpha$ and the associated mometum
$(\Pi_\psi)_\beta$, which, if the Lagrangian $\Lcal$ exists is given by $(\Pi_\psi)_\beta \equiv \partial \Lcal/\partial \dot{\psi}^\beta$ as
\be\label{eq:anticom}
    \{\psi_\alpha, (\Pi_\psi)_\beta\} = i \delta_{\alpha\beta},
\ee
while the (anti)commutator of $\psi_\alpha$ with any operator independent of $(\Pi_\psi)_\alpha$ is taken to vanish. Thus, theories in which $\psi^\dagger_\alpha$ is independent of $(\Pi_\psi)_\alpha$ are problematic because then $\{\psi_\alpha, \psi^\dagger_\beta\} = 0$, which implies $||\psi_\alpha \left|A\right>||^2 = 0$ for any state $\left|A\right>$. This means that either the Hilbert space contains zero-norm states or $\psi_\alpha$ identically vanishes in it.

The only way of avoiding these issues in the free theory via constraints of the form
\be\label{eq:constraint-abstract}
  K_\alpha^\beta (\Pi_\psi)_\beta
  + L_\alpha^\beta \psi^\dagger_\beta
  + M_\alpha^\beta \psi_\beta
  =
  0,
\ee
for some linear (possibly differential, but local) operators $K$, $L$, and $M$. By $\{\psi_\alpha, \psi_\beta\} = 0$ and \eqref{eq:anticom}, this implies that $K_\beta^\gamma \delta_{\alpha\gamma} = i L_\beta^\gamma \{\psi_\alpha, \psi^\dagger_\gamma\}$. Both $K$ and $L$ must be nonsingular in order for the relation to apply to the full set of field operators and their conjugates. As $L$ can be inverted, Eq.~\eqref{eq:constraint-abstract} can be schematically recast as $\{\psi_\alpha, \psi^\dagger_\beta\} = -i (L^{-1}K)_{\alpha\beta}$.
Moreover, if $(L^{-1}K)_{\alpha\beta}$ is a c number, it can be diagonalized because, for a positively definite norm, $\left<A\right|\{\psi_\alpha, \psi^\dagger_\beta\}\left|A\right>$ must be Hermitian and positive for any state $\left|A\right>$. Thus, with a suitable field redefinition it is possible diagonalize $(L^{-1}K)_{\alpha\beta}$. Note that in the first order formulation for spin-1/2, Eq.~\eqref{eq:constraint-abstract} has the form $\{\psi_\alpha, \psi^\dagger_\beta\} \propto \delta_{\alpha\beta}$ which follows from $(\Pi_\psi)_\alpha \equiv \partial \Lcal/\partial \dot{\psi}^\beta \propto \psi_\beta{}^{\dagger}$ and the canonical commutation relations \eqref{eq:anticom}. However, this implies a Lagrangian in which time derivatives appear only through $\dot{\psi}^\beta \psi_\beta{}^{\dagger}$, which, due to Lorentz invariance, is not possible for higher $j$.

In general, for the theory to be Lorentz invariant, the form of Eq.~\eqref{eq:constraint-abstract} should be preserved by Lorentz transformations. In order to provide an explicit expression for $L$ and $K$ when a Lagrangian formulation exists, it is convenient to define
\begin{align}
  \label{eq:3}
  (\widetilde{\Pi}_\psi)^{a\dot{a}}_{a_1 \ldots a_{2j}}
  &\equiv
    \frac{\partial \Lcal}{\partial (\partial_{a \dot{a}} \psi^{a_1 \ldots a_{2j}})},
\end{align}
which depends linearly on $\Pi_\psi$. Then, the Lorentz-invariant expression of the form of Eq.~\eqref{eq:constraint-abstract} containing the lowest number of derivatives is
\be
  (\widetilde{\Pi}_\psi)^{a\dot{a}}_{a_1 a_2 \ldots a_{2j}}
  \propto
  \delta^a_{a_1}
  \,
  \epsilon^{\dot{a} \dot{a}_1}
  \,
  \partial^{a_2 \dot{a}_2} \ldots \partial^{a_{2j} \dot{a}_{2j}}
  \,
  \psi^\dagger_{\dot{a}_1 \ldots \dot{a}_{2j}}.
\ee
Thus, a consistent theory of general-spin fermions seems to require equations of motion with as many as $2j$ derivatives and, in particular, purely second-order quantum theories of fermions with spin-3/2 or higher do not appear to be possible.

\section{Correction to propagators from quadratic terms}
\label{app:quadratic}

\begin{figure}
    \centering
    \includegraphics[width=0.5\textwidth]{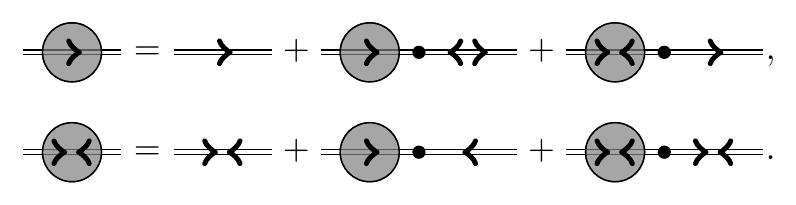}
    \caption{Diagrammatic equations for the resummed propagators. Lines with encircled arrows denote the resummed propagators~\eqref{eq:resummed-propagators} and black dots denote insertions of the quadratic operator~\eqref{eq:quadratic-operator}.}
    \label{fig:resummation}
\end{figure}

Operators quadratic in the general-spin field do not necessarily produce a constant contribution to the mass. To avoid this issue, we assumed that portal coupling \eqref{eq:portal} does not produce quadratic term in the vacuum. The contribution from the quadratic operator,
\be
    - \frac{\lambda v^2}{2} \psi^{(a)}\psi_{(a)} + \hc ,
    \label{eq:quadratic-operator}
\ee
can be easily computed using diagrammatic techniques~\cite{Dreiner:2008tw}. The Feynman rules have a similar structure than the four-legged vertices given in Fig.~\ref{tab:interactions}. The resummed propagators, depicted with encircled arrows in Fig.~\ref{fig:resummation}, can be parametrized as
\be
    \frac{i A(p^2) p_{(a)(\dot{a})} / m^{2j}}{p^2 - m^2},
    \qquad
    \frac{i B(p^2) {\delta_{(a)}}^{(b)} }{p^2 - m^2},
    \label{eq:resummed-propagators}
\ee
and obey the equations shown diagramatically in Fig.~\ref{fig:resummation}. These can be recast as
\bea
    \begin{pmatrix} A \\ B \end{pmatrix}
    =   \begin{pmatrix} 1 \\ 1 \end{pmatrix} + \frac{\vev^2}{p^2 - m^2} 
    \begin{pmatrix} \lambda & \lambda^{*} \\ \lambda (p^2/m^2)^{2j} & \lambda^{*} \end{pmatrix} \begin{pmatrix} A \\ B \end{pmatrix},
\eea
so that, for example, the propagator with an arrow pointing to the right takes the form
\be
    \frac{i p_{(a)(\dot{a})} / m^{2j}}{p^2 - m^2 - 2\vev^2 \re{\lambda} - \frac{\vev^4|\lambda|^2}{m^2} \sum^{2j-1}_{n=0}(p^2/m^2)^{n} }.
\ee
The other two propagators are resummed analogously.

The corrections can be absorbed by a redefinition of the pole mass only in the lower spin cases: 
\begin{itemize}
    \item For $j = 0$, the squared pole mass is
$$
    m^2 + 2\vev^2 \re{\lambda},
$$
as expected for a scalar.
    \item For $j = 1/2$, the squared pole mass is 
$$
   m^2 |1 + \lambda \vev^2/m^2  |^2.
$$
Additionally, in order to recover the correct propagator normalization $p_{(a)(\dot{a})} / m_{\rm pole}^{2j}$, the field must be rescaled as $\psi \to \psi /\sqrt{1 + \lambda \vev^2/m^2}$ affecting all interactions with $\psi$.
    \item For $j = 1$, the squared pole mass is 
$$
    m^2 \frac{|1 + \lambda \vev^2/m^2 |^2}{1 - \left|\lambda \vev^2/m^2 \right|^2},
$$
    and the field must be rescaled, as $\psi \to \psi \frac{1 - \left|\lambda \vev^2/m^2 \right|^2}{|1+\lambda \vev^2/m^2|}$.
    \item For $j \geq 1$, the squared pole mass can be estimated as
$$
    m^2 |1 + \lambda \vev^2/m^2 |^2 + (2j-1)|\lambda \vev^2/m |^2 + \mathcal{O}(\lambda^3),
$$
assuming $\lambda \vev^2 \ll m^2$. The resummed propagator denominator contains powers up to  $(p^2/m^2)^{2j - 1}$, implying that there are nonperturbative poles scaling as $\lambda^{-1/(2j-1)}$. These, however, do not show up in the EFT approach as such a resummation is not justified.
\end{itemize}
In the EFT description, the corrections to the propagator must be subdominant: thus, one must impose
\be
    |\lambda \vev^2/m^2|^2(p^2/m^2)^{2j - 1} \ll 1,
\ee
when $j \geq 1$. In terms of the effective cutoff scale \eqref{def:Lambda}, this condition reads
\be
    p^2 \ll \Lambdaunit^2 \left[  \frac{m\Lambdaunit}{4\pi \vev^2} \right]^{\frac{2}{2j-1}}.
\ee

\begin{figure*}
    \includegraphics[width=0.93\textwidth]{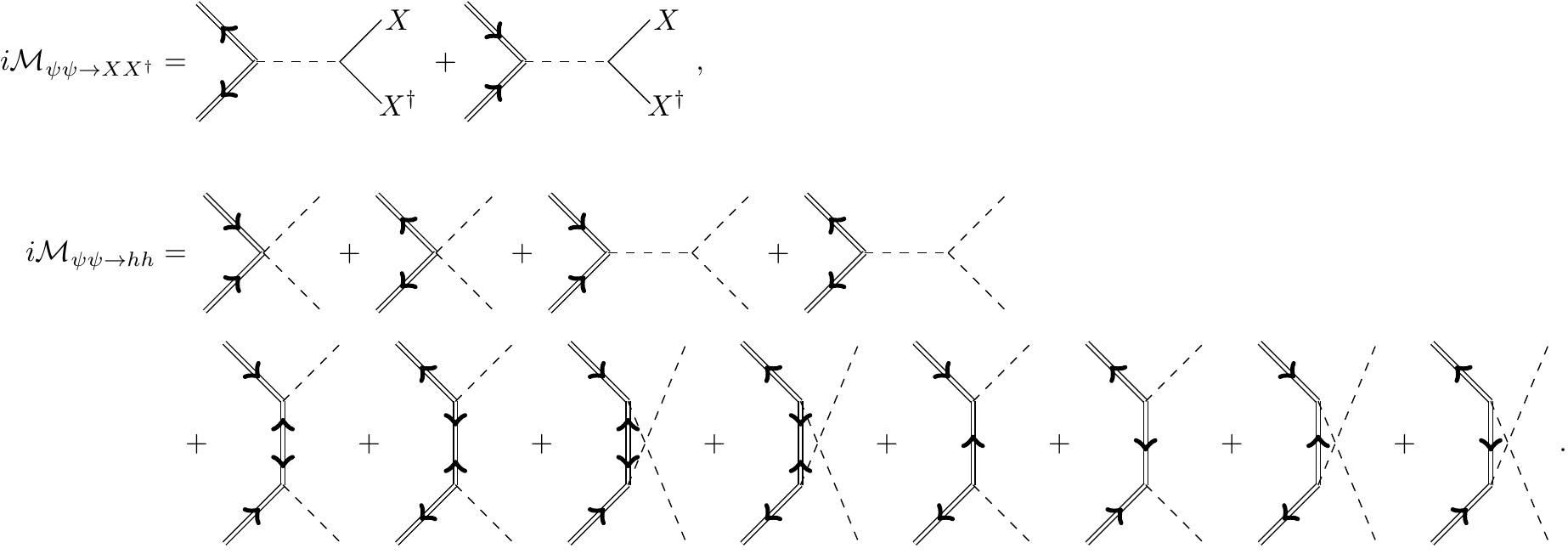} \\
    \caption{Diagrams contributing to the annihilation of two $\psi$ particles into two SM ones. Dashed lines represent the Higgs boson $h$. The label $X$ corresponds either to a massive SM fermion or to a massive SM gauge boson.}
    \label{fig:annihilation}
\end{figure*}

\section{Spin-3/2 alternatives}
\label{app:spin-32}

Here we consider alternative spin-3/2 theories in which either the Lorentz-group irrep or the SM-gauge-group irrep differ from the ones we have considered so far, and write down the leading-order effective Hamiltonian corresponding to each case.\footnote{To obtain a list of all the independent operators allowed at each order, we have used the code \texttt{BasisGen}~\cite{Criado:2019ugp}.} This serves as an example of what may happen for other spins when one goes beyond the minimal case. In general, variations in the Lorentz group irrep should not change the physics, as long as only the spin-$j$ degrees of are coupled (as discussed in Sec.~\ref{sec:EFT}). On the other hand, different gauge-group irreps give rise to physically different possibilities.

\subsection{$\R_{3/2}$}

In this case, the field $\psi$ must belong to a real representation of the SM gauge group. This means that it should have vanishing hypercharge, integer electroweak isospin and zero $SU(3)$ triality. The simplest case corresponds to having an $(1, 1)_0$ SM irrep, which is the one whose phenomenology is studied in detail in Sec.~\ref{sec:pheno}. The next simplest case is given by the $(1, 3)_0$ irrep, for which the most general $\mathbb{Z}_2$-symmetric interaction Hamiltonian for SM-$\psi$ interactions at the lowest order in $1/\Lambda$ is
\begin{align}
  \Hcal_{\text{int},\text{SM}}
  &=
    - \lambda |\phi|^2 \psi_{abc} \psi^{abc}
    \nonumber
  \\
  &\phantom{=}
    + c_W \epsilon_{ABC} W^A_{ab} \psi^B_{acd} \psi^{C\,bcd}
    + \hc,
\end{align}
where upper case latin letters $A$, $B$ denote $SU(2)$ triplet indices and ${(W^A)_a}^b = {(\sigma^{\mu\nu})_a}^b (W^A_{\mu\nu} + i\widetilde{W}^A_{\mu\nu})/2$ with $W^A_{\mu\nu}$ the $SU(2)$ field-strength tensor. 

Here and below, additional interactions with gauge fields are implied by the minimal substitution of partial derivatives with gauge covariant ones in the eom \eqref{eq:eom} and \eqref{eq:klein-gordon}. Such interactions are not listed here.

\subsection{Charged $(3/2, 0) \oplus (0, 3/2)$}

In order for the field not to reduce to two independent copies of the $\R_{3/2}$ representation, it must belong to a complex representation of the SM gauge group. The simplest such representation containing a neutral component (to be used as the DM candidate) is an $SU(2)$ doublet with hypercharge $Y = 1/2$. The $\mathbb{Z}_2$-symmetric leading-order interactions with the SM are given by
\begin{align}
  \Hcal_{\text{int},\text{SM}}
  &=
    - \lambda_L |\phi|^2 \psi_L^2
    - \lambda_R |\phi|^2 \psi_R^2
  \nonumber \\
  &\phantom{=}
    - \lambda_{LR1} |\phi|^2 \psi_{L\,abc} \psi^{\dagger\,abc}_R
  \nonumber \\
  &\phantom{=}
    - \lambda_{LR2} (\phi^\dagger \psi_{L\,abc}) (\psi^{\dagger\,abc}_R \phi)
  \nonumber \\
  &\phantom{=}
    + c_W W^A_{ab} \psi^{\dagger\,B}_{R\,acd} \sigma^A \psi^{C\,bcd}_L
  \nonumber \\
  &\phantom{=}
    + c_B B_{ab} \psi^{\dagger\,B}_{R\,acd} \psi^{C\,bcd}_L
    + \hc,
\end{align}
where ${B_a}^b = {(\sigma^{\mu\nu})_a}^b (B_{\mu\nu} + i\widetilde{B}_{\mu\nu})/2$ with $B_{\mu\nu}$ the $U(1)$ field-strength tensor.

\subsection{Neutral $(1, 1/2) \oplus (1/2, 1)$}

The field irrep here is defined as the subset of the $(1, 1/2) \oplus (1/2, 1)$ representation satisfying $\psi \equiv \psi_L = \psi_R^\dagger$, where $\psi_L$ and $\psi_R$ belong to the $(1, 1/2)$ and $(1/2, 1)$ sectors, respectively. The conditions over the SM irrep are the same as in the $(3/2, 0)$ case. We assume that the usual Rarita-Schwinger formulation is used, so the field has dimension $3/2$. Then, the leading-order interaction terms have dimension 5. The full interaction Hamiltonians for the $(1, 1)_0$ and $(1, 3)_0$ cases at this order are:
\be
  \Hcal_{\text{int}}
  =
  -\frac{\lambda}{\Lambda} \psi_{ab\dot{c}} \psi^{ab\dot{c}} |\phi|^2
  + \frac{(c_l)_i}{\Lambda} \psi_{ab\dot{c}} l_i^{a} D^{b\dot{c}} \phi
  + \hc,
\ee
and 
\begin{align}
  \Hcal_{\text{int}}
  &=
    -\frac{\lambda}{\Lambda} \psi^A_{ab\dot{c}} \psi^{A\,ab\dot{c}} |\phi|^2
    + \frac{(c_l)_i}{\Lambda} \psi^A_{ab\dot{c}} (l_i^{a})^T \sigma^A D^{b\dot{c}} \phi
    \nonumber \\
  &\phantom{=}
    + \frac{c_{B1}}{\Lambda} {B_d}^a \, \psi^A_{ab\dot{c}} \psi^{A\,db\dot{c}}
    + \frac{c_{B2}}{\Lambda} {(B^\dagger)_{\dot{d}}}^{\dot{c}} \, \psi^A_{ab\dot{c}} \psi^{A\,ab\dot{d}}
    \nonumber \\
  &\phantom{=}
    + \hc ,
\end{align}
where $l_i$ is the $i$th generation SM lepton doublet and ${}^T$ denotes $SU(2)$ doublet transposition. Self interactions are not allowed at this order: they would appear at dimension 6.

Since the field we are using now contains unphysical degrees of freedom, a careful examination of the interaction Hamiltonian is needed to see which conditions need to be applied to it, in order for the unphysical components not to be coupled with the physical ones.

\subsection{Charged $(1, 1/2) \oplus (1/2, 1)$}

As for the charged $(3/2, 0) \oplus (0, 3/2)$ field, the simplest SM irrep here is an $SU(2)$ doublet with hypercharge $Y = 1/2$, and as for the neutral $(1, 1/2) \oplus (1/2, 1)$, we assume that the dimension of the field is $3/2$. Then, we have
\begin{align}
  \Hcal_{\text{int}}
  &=
    \frac{c_{B1}}{\Lambda} {B^a}_d \, \psi^\dagger_{R\,ab\dot{c}} \psi_L^{db\dot{c}}
    \nonumber
    \\
  &\phantom{=}
    + \frac{c_{B2}}{\Lambda} {(B^\dagger)^{\dot{c}}}_{\dot{d}}
    \, \psi^\dagger_{R\,ab\dot{c}} \psi_L^{ab\dot{d}}
    \nonumber
    \\
  &\phantom{=}
    + \frac{c_{W1}}{\Lambda} {(W^A)^a}_d \, \psi^\dagger_{R,ab\dot{c}} \sigma^A \psi_L^{db\dot{c}}
    \nonumber
\end{align}
\begin{align}  
  &\phantom{=}
    + \frac{c_{W2}}{\Lambda} {(W^{\dagger A})^{\dot{c}}}_{\dot{d}}
    \, \psi^\dagger_{R\,ab\dot{c}} \sigma^A \psi_L^{ab\dot{d}}
    \nonumber
  \\
  &\phantom{=}
    + \frac{(c_{Bl})_i}{\Lambda} (B^\dagger)^{\dot{a}\dot{c}} \, \psi_{R\,\dot{a}\dot{b}c} l^c_i
    \nonumber
  \\
  &\phantom{=}
    + \frac{(c_{Wl})_i}{\Lambda} (W^{\dagger A})^{\dot{a}\dot{c}} \, \psi_{R\,\dot{a}\dot{b}c} \sigma^A l^c_i
    \nonumber
    \\
  &\phantom{=}
    + \frac{(c_e)_i}{\Lambda} \psi_{R\,ab\dot{c}} e_i^{a} D^{b\dot{c}} \phi
    + \frac{c_\phi}{\Lambda} \psi^\dagger_{R\,ab\dot{c}} \psi_L^{ab\dot{c}} |\phi|^2
    \nonumber
  \\
  &\phantom{=}
    - \frac{\lambda_{L}}{\Lambda}
    (\phi^\dagger \psi_{L\,ab\dot{c}}) (\phi^\dagger \psi_L^{ab\dot{c}})
    \nonumber
  \\
  &\phantom{=}
    - \frac{\lambda_{R}}{\Lambda}
    (\phi^\dagger \psi_{R\,\dot{a}\dot{b}c}) (\phi^\dagger \psi_R^{\dot{a}\dot{b}c})
    + \hc .
\end{align}
Restrictions on the structure of $\Hcal_{\text{int}}$ have to be imposed to decouple the unphysical components of the field, as for its neutral counterpart.

\vfill

\onecolumngrid
    \section{Annihilation cross section}
    \label{app:xs-formulas}
    
    The diagrams contributing to the annihilation of two $\psi$ particles into a pair of SM model ones are shown in Fig.~\ref{fig:annihilation}. The cross section for annihilation into massive SM particles (except the Higgs boson) is
    \be
        \sigma_{\psi\psi \to X X^\dagger}
        =   
        \frac{\vev^2 |\lambda|^2}{2 \pi (2j + 1)^2 s (s - m_h^2)^2}
        \sqrt{\frac{s - 4m_X^2}{s - 4m^2}}
        \brackets{
            \frac{(p_1 p_2)^{~(a)}_{(a)}}{m^{4j}}+\cos 2\theta (-1)^{2j}\delta^{(a)}_{(a)}
        }
        \gamma _X(s),
    \ee
    where $p_1$ and $p_2$ are the four momenta of the incoming particles and, $s$ is the center-of-mass energy squared, and
    \be
        \gamma_f(s) = \frac{2 m_f^2}{\vev^2} \parens{s - 4 m_f^2}, 
        \qquad
        \gamma_W(s) = \frac{4 m_W^4}{\vev^2} \parens{
            \frac{s^2}{4 m_W^4} - \frac{s}{m_W^2} + 3
        },
        \qquad
        \gamma_Z(s) = \frac{2 m_Z^4}{\vev^2} \parens{
            \frac{s^2}{4 m_Z^4} - \frac{s}{m_Z^2} + 3
        },
    \ee
    with $f$ representing any massive SM fermion. We omitted explicit indices in the multispinor traces for brevity, e.g. $(p_1 p_2)^{(a)}_{(a)} \equiv p_1{}_{(a)(\dot a)} p_2^{{(\dot a)(a)}}$. All multispinor traces encountered here can be computed using Eqs.~\eqref{eq:prod_sym} and~\eqref{eq:trace_sym}.
    
    The differential cross section for the annihilation into two Higgs bosons, $\psi(p_1)\psi(p_2)\to h(k_1)h(k_2)$, is given by
    %
    %
    \bea
   \frac{d\sigma_{\psi\psi \to hh}}{dt}
        =
        \frac{1}{32 (2j + 1)^2 \pi s (s - 4m^2)}
  &\Bigg\{
2|F|^2 \frac{(p_1p_2)^{~(a)}_{(a)}}{m^{4j}} +2(-1)^{2j}\delta^{(a)}_{(a)}\re{F^2}\\
+\frac{16\re{F}v^2 |\lambda|^2}{m^{4j}}
&   \Bigg[
\frac{(p_{1}(p_1-k_1))_{(a)}^{~(a)}+(-1)^{2j}(p_{2}(p_1-k_1))_{(a)}^{~(a)}}{t-m^2}\\
&   +\frac{(p_{1}(p_1-k_2)_{(a)}^{~(a)}+(-1)^{2j}(p_{2}(p_1-k_2))_{(a)}^{~(a)}}{u-m^2}
\Bigg]\\
\frac{32v^4 |\lambda|^4}{m^{8j}}
&   \Bigg[
\frac{(p_{1}(p_1-k_1) p_{2}(p_1-k_1))_{(a)}^{~(a)}
+ (2j+1)(-t m^{2})^{2j}}{(t-m^2)^2}\\
&   +\frac{(p_{1}(p_1-k_2) p_{2}(p_1-k_2))_{(a)}^{~(a)}
+(2j+1)(-um^2)^{2j}}{(u-m^2)^2}\\
  +\frac{1}{(t-m^2)(u-m^2)}&\Big(
(p_{1}(p_1-k_1) p_{2}(p_1-k_2))_{(a)}^{~(a)}
   +(p_{1}(p_1-k_2) p_{2}(p_1-k_1))_{(a)}^{~(a)}\\
&+2(-1)^{2j}((p_1-k_1)(p_1-k_2))_{(a)}^{~(a)}m^{4j}\Big)
\Bigg]\Bigg\},
\eea
    where
    \begin{align}
        F &= 2 \lambda \parens{
        1 + \frac{3 m_h^2}{s - m_h^2}
        - \frac{2 \vev^2 \lambda}{t - m^2}
        - \frac{2 \vev^2 \lambda}{u - m^2}
        },
    \end{align}
    with $s$, $t$ and $u$ being the usual Mandelstam variables.
    As $s \to 4 m^2$,
    \begin{align}
        \sigma_{\psi\psi \to \bar{f}f} \vel
        &\sim
        \frac{
            2 |\lambda|^2 m_f^2 (m^2 - m_f^2)^{3/2}
        }{
            (2j + 1) \pi m^3 (4 m^2 - m_h^2)^2
        }
        \Big[1 + (-1)^{2j} c_{2\theta} + \frac{2}{3}j(j+1)\vel^2\Big],
        \\
        \sigma_{\psi\psi \to VV} \vel
        &\sim
        \frac{
             \eta_V |\lambda|^2 m_V^4 (m^2 - m_V^2)^{1/2}
            \parens{4 \frac{m^4}{m_V^4} - 4 \frac{m^2}{m_V^2} + 3}
        }{
            (2j + 1)  \pi m^3 (4m^2 - m^2_h)^2
        }
        \Big[1 + (-1)^{2j} c_{2\theta} + \frac{2}{3}j(j+1)\vel^2\Big],
    \end{align}
    where, as in Eq.~\eqref{eq:non-rel-sigma}, we assumed $c_{2\theta} = (-1)^{2j}$ in the $\vel^2$ term.
    
\twocolumngrid

\bibliography{any_spin}

\end{document}